\documentclass[aps,prd,preprint,onecolumn,amsmath,amssymb,floatfix]{revtex4-2}
\usepackage{graphicx}
\usepackage{rotating}
\usepackage{color}
\usepackage{soul}
\usepackage{hyperref}
\usepackage{threeparttable}
\usepackage{natbib} 
\usepackage{longtable}
\usepackage{appendix}
\usepackage{array}   
\usepackage{tabularx}
\usepackage[utf8]{inputenc}

\newcommand{\bfp}{\mbox{\boldmath $p$}}

\def\be{\begin{equation}}
\def\ee{\end{equation}}
\def\ba{\begin{eqnarray}}
\def\ea{\end{eqnarray}}
\def\bl#1\el{\begin{align}#1\end{align}}

\def\l{\left}
\def\r{\right}

\begin{document}
\title{Strong magnetic field inside degenerate relativistic plasma and the impacts on the neutrino transport in Core-Collapse Supernovae}

\author{Y. Luo$^{1}$}
\email{Corresponding author: yudong.luo@pku.edu.cn}
\author{S. Zha$^{2}$}
\email{zhashuai@ynao.ac.cn}
\author{T. Kajino$^{3,4,5}$}
\email{kajino@buaa.edu.cn}

\affiliation{School of Physics and Kavli Institute for Astronomy and Astrophysics, Peking University, Beijing 100871, China}
\affiliation{International Centre of Supernovae (ICESUN), Yunnan Key Laboratory of Supernova Research, Yunnan Observatories, Chinese Academy of Sciences (CAS), Kunming 650216, People's Republic of China}
\affiliation{School of Physics, Peng Huanwu Collaborative Center for Research and Education, and International Research Center for Big-Bang Cosmology and Element Genesis, Beihang University, Beijing 100191, China}
\affiliation{Graduate School of Science, The University of Tokyo, Tokyo 113-0033, Japan}
\affiliation{National Astronomical Observatory of Japan, Tokyo 181-8588, Japan}

\date{\today}
\begin{abstract}
We investigate the impacts of strong magnetic fields on neutrino transport in core-collapse supernovae (CCSNe) using the leakage scheme. Magnetic field quantizes the momentum of electrons and positrons, resulting in the modification of weak-interaction cross sections and the chemical potentials of electrons and positrons. We derive formula for neutrino leakage scheme including these two impacts and perform 1D CCSN simulations with \texttt{GR1D}. Magnetic field strengths from $10^{16}$\,G to $10^{17}$\,G were applied to postbounce phase. The results show that the neutrino opacities are enhanced due to the amplified interaction rates, with stronger effects on anti-neutrinos. This leads to larger neutrinosphere radii, longer neutrino trapping timescales, reduced peak luminosities, and delayed peak energies. 
\end{abstract}

\maketitle
\section{Introduction}
Core-collapse supernovae (CCSNe) are one of the major astronomical neutrino sources since the collapse of the iron core of massive stars releases gravitational energy of $\sim10^{53}$ ergs and neutrinos carry away most of this energy \citep{Bethe:1990mw}. This explosive scenario has been confirmed by the detection of about two dozen neutrino events from SN1987A occurring in the Large Magellanic Cloud, with several Cherenkov detectors, including Kamiokande \citep{Kamiokande-II:1987idp}, Irvine–Michigan–Brookhaven detectors \citep{Bionta:1987qt}, and the Mont Blanc Underground Neutrino Observatory \citep{Aglietta:1987it}. 

Neutrinos are both essential in the explosion dynamics and nucleosynthesis. The competition between neutrino heating and accretion onto the proto-neutron star (PNS) decides if the shock could successful revival. 
However, neutrino transport from opaque to transparent regions is among the most crucial and challenging parts to model CCSN, owing to the large variety of scales involved and the fine spatial resolutions needed to obtain accurate results. One efficient way to model the neutrino transport is the leakage scheme \citep{1981ApJ...249..270V,Bruenn:1985en,2000ApJ...539..865B,1996A&A...306..167J,Rosswog:2003rv,Ruffert:1995fs}, it describes the neutrino spectrum as a Fermi-Dirac distribution with a local chemical potential $\mu_{\nu_{i}}$ for a neutrino species $i$. Neutrinos are approximately treated as they leak from the neutrino-sphere $R_{\nu_i}$, which by definition is where the neutrino optical depth $\tau_{\nu_i}$ equals to $2/3$. On the other hand, the neutrino transport during the post-bounce phase determines the electron fraction ($Y_e$) of the ejecta, which is the key parameter for nucleosynthesis in the later phase, especially for the rapid neutron-capture process \citep{1994ApJ...433..229W,Nishimura:2015nca}. The emitted neutrinos from the neutrinosphere can further affect nucleosynthesis via the neutrino-nucleus reactions, i.e., $\nu$-process and $\nu $p-process \citep{Woosley:1989bd,Frohlich:2005ys,Wanajo:2006ec}. The neutrino luminosities and spectra at the neutrinosphere are critical inputs for the nucleosynthesis calculation, and those parameters are obtained from the neutrino transport during explosion \citep{Wanajo:2010mc,Balasi:2015dba}. 

On the other hand, {numerical studies have suggested that the magnetic field especially the toroidal component near the newborn PNS surface the magnetic field of a newborn PNS could possibly reach values up to $10^{15-17}$ G \citep{nakamura15, kiuchi15, kiuchi14,takiwaki09,price,ruiz,Mosta:2015ucs,Raynaud:2020ist,Obergaulinger:2020cqq,Mosta:2014jaa,powell23}
while the neutrino transport inside such a strong magnetic field is not well studied.} The neutrino transport could be significantly altered due to the modification of weak interactions by the magnetic field, while few studies focused on the impacts of magnetic field on microscopic process and the accompanying impacts on astrophysical sites: \cite{1991ApJ...383..745L,Chakrabarty:1996te, Chakrabarty:1997ef} studied the equation of state (EOS) under a strong magnetic field, \cite{Famiano:2020fbq} calculated the $r$-process nucleosynthesis considering impacts of a strong magnetic field on beta decay and electron-positron capture rates, and \cite{Maruyama:2021ghf,Famiano:2022lmz} calculated the neutron star cooling under a strong magnetic field. Several previous studies investigated the neutrino-nucleon interactions inside the magnetic field of PNS \citep{Duan:2004nc,Lai:1998sz,Arras:1998mv} while they did not investigated the effects in CCSN dynamical simulations. Recently, the magnetic field in M1 neutrino transport scheme has been explored in both 3D Magnetohydrodynamical (MHD) CCSN simulation \citet{2021ApJ...906..128K} and 
1-D simulation for various magnetic field configuration \cite{luo25}. For complementary purposes, in this study, we developed a formula that integrates the magnetic field effect with the leakage scheme. {We set the magnetic field strengths from $10^{16}$\,G to $10^{17}$\,G. Our purpose is to explore how neutrino transport is modified in this strongly magnetized limit. We also emphasize that the present study does not include magnetic-pressure feedback on the hydrodynamics. Instead, we isolate the magnetic-field correction to neutrino transport as an additional physical effect that can be incorporated into future multidimensional MHD simulations.}

Compared to the M1 scheme used in previous studies, this method provides an effective way to estimate crucial physical properties of neutrinos, such as luminosity, neutrinosphere radius, and mean neutrino energy.
This paper is arranged as follows. We show the impacts of magnetic field on weak interactions in Section \ref{B_impact}. In Section \ref{nu_trans}, we first describe the leakage scheme for neutrino transport and the corresponding adaption for the presence of a strong magnetic field. Then, we present the numerical results of CCSN simulations and the impact of magnetic field strength on neutrino signals. In Section \ref{con_dis}, we summarize the results. 
\section{Weak Interactions in Strong Magnetic Field}\label{B_impact}
Similar to our previous study on M1 neutrino transport scheme \cite{luo25}, we apply the cross section from \cite{Duan:2004nc,Duan:2005fc} with the $0$-th order of approximation:
\begin{eqnarray}
\label{sigma_B}
   \sigma_{\rm \nu N}(E_n,B) &=& \sigma_B^1\Big[1+2\chi \dfrac{(f\pm g)g }{ f^2 +3g^2} \cos \Theta_\nu\Big]+\nonumber  \\
   & \sigma_B^2&\Big[\dfrac{f^2 -g^2 }{ f^2 +3g^2}\cos\Theta_\nu + 2\chi \dfrac{(f\mp g) g }{ f^2 + 3g^2} \Big],
\end{eqnarray}
where
\begin{eqnarray}
   &&\sigma_B^1 = \dfrac{G_F^2 \cos^2\theta_C }{ 2\pi}(f^2 + 3g^2)\times \nonumber\\
   &&eB\sum^{n_{\rm max}}_{n=0}\sum^{s=1}_{s=-1}\dfrac{g_nE_n }{ \sqrt{E_n^2 - m_e^2 -(2n+s-1)eB}},
\end{eqnarray}
and
\begin{equation}
\sigma_B^2 = \dfrac{G_F^2 \cos^2\theta_C }{ 2\pi}(f^2 + 3g^2)eB\dfrac{E_n}{ \sqrt{E_n^2 - m_e^2}}. 
\end{equation}
where $G_F=1.166\times 10^{-5}$GeV$^{-2}$ is the Fermi coupling constant, $\theta_C$ is the Cabbibo angle, for which we set the value $\rm cos^2(\theta_C)=0.95$. $f = 1$ and $g = 1.26$ are the vector and axial-vector coupling coefficients, respectively. $E_n$ is the quantized electron energy $ E_n^2 = p_z^2+m_e^2+ (2n+s-1)eB$ with $n$ is the Landau quantum number \citep{Landau:1991wop} and $s$ stands for the spin of $e^\pm$, i.e., $s=1$ ($s=-1$) corresponds to the spin (anti)parallel to the magnetic field. In Eq. \ref{sigma_B}, the upper sign is for $\nu_e +n$ reaction, and the lower sign is for $\bar{\nu}_e + p$. $\Theta_\nu $ is the angle between neutrino momentum and the magnetic field, and as we will show, this angular-dependence cross section can be neglected and we set $\Theta_\nu=0$ through this work.

Inside the magnetic field, $e^\pm$ obey the Fermi-Dirac distribution but with the Landau quantization of $E_n$ so that the chemical potential $\mu_e$ also need a correction. $\mu_e$ is evaluated from the charge neutrality of plasma, i.e., the net number density of both electrons and positrons should be balanced with the ions, where $N_A$ is the Avogadro number. For a magnetized plasma, the momentum space of $e^\pm$ is quantized as
\begin{eqnarray}
&&2\dfrac{d^3p}{ (2\pi)^3} f_{\rm FD}(E_e,\pm\mu_e;T)\nonumber\\
    \nonumber\\
&&\to eB\sum^{\infty}_{n=0} (2-\delta_{n0}) \dfrac{dp_z}{ 4\pi^2}f_{\rm FD}(E_n,\pm\mu_e;T),
\end{eqnarray}
then, the net electron number density is given by
\begin{eqnarray}
\label{chemi_po}
   \rho N_A Y_e = &n_e&= \frac{m_e\omega_c}{(2\pi)^2}\sum_0^{n_{\rm max}} \int d\bfp \Big[f_{\rm FD}(E_n, +\mu_e; T)_{s=-1}\nonumber\\
   &+&f_{\rm FD}(E_n, + \mu_e; T)_{s=1}
        - f_{\rm FD}(E_n, -\mu_e; T)_{s=-1}\nonumber\\
        &-&f_{\rm FD}(E_n, -\mu_e; T)_{s=1}\Big].
\end{eqnarray}
Here, $\omega_c = eB/m_e$ is the cyclotron frequency, and $s=\pm1$ represents the spin state of $e^\pm$. Notice that inside the magnetic field, the spin is only parallel or anti-parallel with the magnetic field. 
\begin{figure}[h!]
\centering
\includegraphics[width=0.5\textwidth,clip=true,trim=0cm 0cm 0cm 0cm]{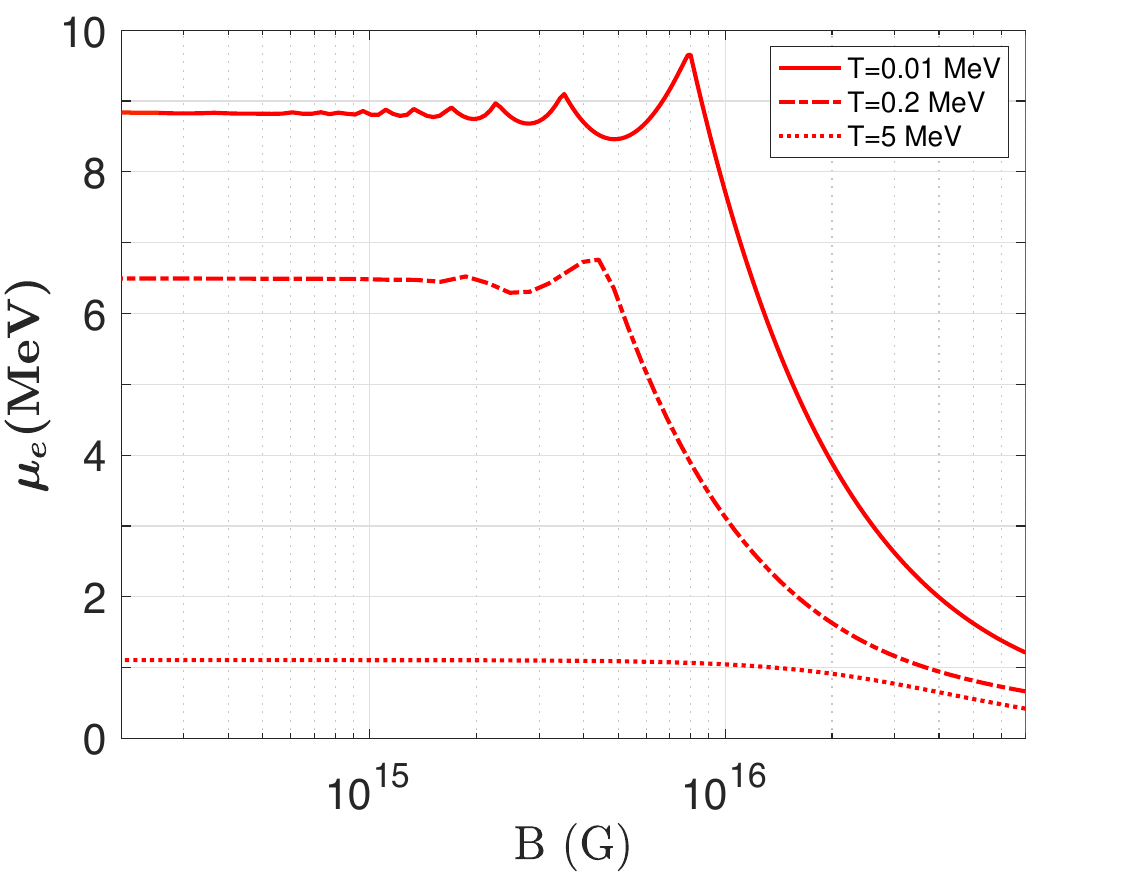}
\includegraphics[width=0.5\textwidth,clip=true,trim=0cm 0cm 0cm 0cm]{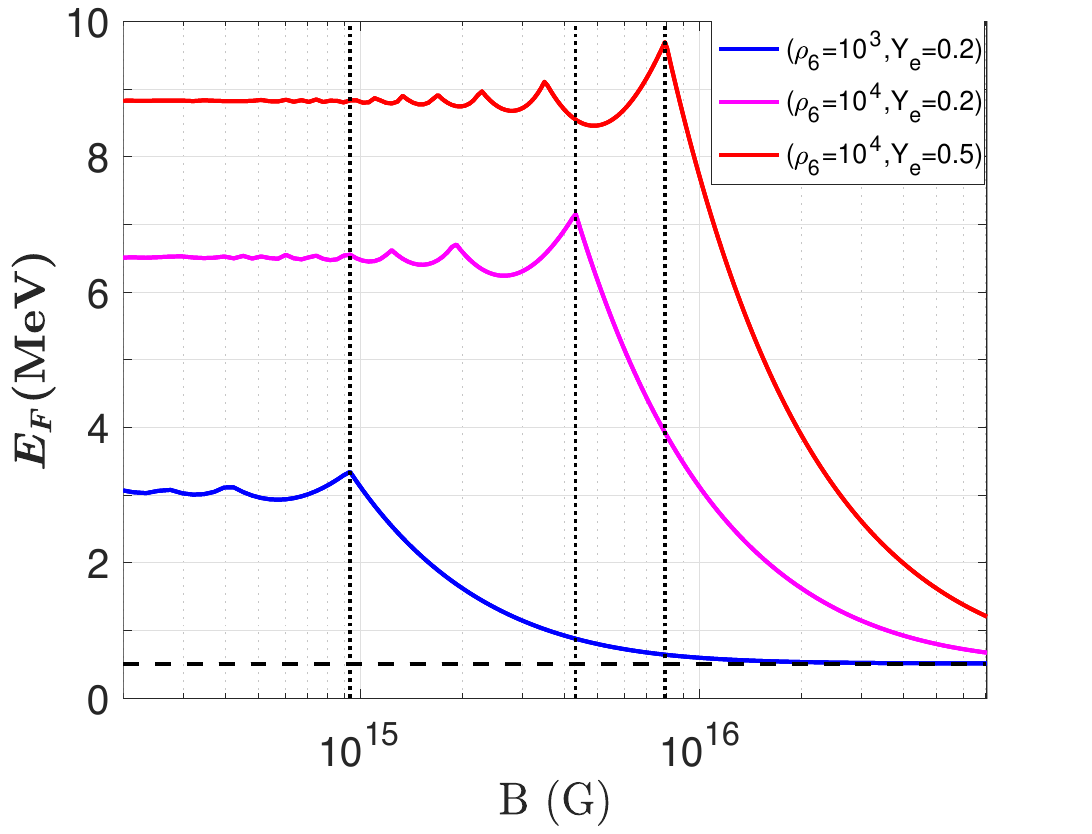}
\caption{\label{Chemi_relation} Upper panel: The chemical potential as a function of magnetic field strength for different temperatures, with $\rho = 10^{10}\rm \ g\ cm^{-3}$and $Y_e = 0.5$. The solid line, dashed line and dotted line correspond to $T=0.01\ \rm MeV$, $T=0.2\ \rm MeV$ and $T=5\ \rm MeV$, respectively. The solid line is consistent with the red line on the lower panel since under this density and temperature. electrons are already in a degenerate state. Lower panel: Fermi energy $E_F$ as a function of magnetic field strength with $\rho_6, Y_e$ for different combinations as $(10^3, 0.2)$ $(10^4, 0.2)$ and $(10^4, 0.5)$, respectively. If the magnetic field is relatively weak, $E_F$ becomes the same as a field-free case. On the other hand, $E_F$ approaches $m_e=0.511$ MeV (horizontal dashed line) when $B>B_{\rm crit}$, vertical dotted lines represent the value of $B_{\rm crit}$ in each set of $\rho$ and $T$.}
\end{figure}

The upper panel of Fig.\ref{Chemi_relation} shows chemical potential as a function of magnetic field strength at different temperatures. We set $\rho_6\equiv \rho/10^6 \rm g\ cm^{-3} = 10^4$ and $Y_e=0.5$ in this panel. In general, chemical potential increases as temperature decreases (from dotted line to solid line in the upper panel of Fig. \ref{Chemi_relation}). While for an extremely low temperature, Eq. \ref{chemi_po} will recover the degenerate case, i.e., the red solid lines in both panels are identical to each other (temperature is set to be zero in the bottom panel of Fig. \ref{Chemi_relation}). 

On the other hand, the chemical potential is reduced with increasing magnetic field strength. This is due to the extra energy contributed from the field to $e^\pm$. In the case of strong fields, $e^\pm$ only occupy the lowest Landau level (LLL), so that $\mu_e$ decreases monotonically as a function of $B$, and approaches the asymptotic value $m_e$ for an extremely strong magnetic field. In the case of degenerate plasma, the critical magnetic field strength $B_{\rm crit}$ for the scenario that only LLL is occupied reads \cite{DELSANTE1980135}
\begin{equation}
   B_{\rm crit}=\frac{\pi}{e}\Big[2\pi(\rho Y_e)^2\Big]^{1/3}.
\end{equation}
 When the magnetic field strength is below $B_{\rm crit}$, $E_F$ shows a zig-zag pattern (bottom panel of Fig.\ref{Chemi_relation}). 
{The reason is as follows. For an extremely strong magnetic field, only the lowest Landau level (LLL) is occupied. As the magnetic-field strength decreases, higher Landau levels become accessible one after another. Whenever a new Landau level starts to be populated, the available phase space increases, and the Fermi energy $E_F$ decreases accordingly for a fixed electron number density. Between two such thresholds, this readjustment becomes weaker and $E_F$ increases again more smoothly. This pattern is repeated whenever the next Landau level appears, leading to the oscillation-like behavior. We note that this behavior is consistent with Ref.~\cite{DELSANTE1980135}, where only the zero-temperature Fermi energy $E_F$ was discussed. In the present work, we extend their formulation to the finite-temperature case.
}
For a weak magnetic field ($B<10^{15}$ G), $E_F$ values recover to the non-magnetized plasma cases: $E_F=3.01$ MeV (blue line), $6.51$ MeV (pink line), and $8.82$ MeV (red line) for the $(\rho_6, Y_e)$ combinations we chose.

\section{The neutrino transport inside magnetic field}\label{nu_trans}
\subsection{Method}
In leakage scheme, one can calculate the optical depth from the integration of local opacity $\kappa_{ \rm tot}$, which is determined by the local mean free path of neutrinos or, equivalently, by the interaction rates of the neutrino processes (scattering, absorption, and emission). The total opacity $\kappa_{\rm tot}$ of electron type neutrinos are given by \citep{Ruffert:1995fs}
\begin{eqnarray}
   &&\kappa_{{\rm  tot},j}(\nu_e) =  \kappa_{a,j}(\nu_e n)+ \kappa_{s,j}(\nu_e n) + \kappa_{s,j}(\nu_e p) \\
   \nonumber\\
    &&\kappa_{{\rm  tot},j}(\bar{\nu}_e) =  \kappa_{a,j}(\bar{\nu}_e p)+ \kappa_{s,j}(\bar{\nu}_e n) + \kappa_{s,j}(\bar{\nu}_e p),
\end{eqnarray}
where $a$ represents the absorption process and $s$ stands for scattering process. Index $j$ is introduced to denote the opacities for neutrino-number transport ($j=0$) and neutrino-energy transport ($j=1$), respectively. As discussed in Sec. \ref{B_impact}, inside a magnetic field, the absorption cross sections differ from the field-free case and the absorption opacities of ${\nu}_e$ and $\bar{\nu}_e$ read
\begin{eqnarray}
\label{k_abs1}
&&\kappa^B_{a,j}(\nu_e n)=A\rho Y_{\rm np}\times\\
&&\frac{
      \int_0^\infty dE_{\nu_e}\sigma(E_n,B)E_{\nu_e}^{2+j} f_{\rm FD}(E_{\nu_e},\eta_{\nu_e}; T_{\nu_e}) g\Big[E_n,\mu_e(B);T_e\Big]
      }%
  {\int_0^\infty dE_{\nu_e} E_{\nu_e}^{2+j} f_{\rm FD}(E_{\nu_e},\eta_{\nu_e}; T_{\nu_e})},\nonumber\\
 \nonumber\\
 \label{k_abs2}
&&\kappa^B_{a,j}(\bar{\nu}_e p) =    A\rho Y_{\rm pn}\times\\
&&\frac{\begin{aligned}\int_0^\infty dE_{\bar{\nu}_e}\sigma(E_{n},B)E_{\bar{\nu}_e}^{2+j} f_{\rm FD}(E_{\bar{\nu}_e},\eta_{\bar{\nu}_e}; T_{\bar{\nu}_e}) g\Big[E_n,-\mu_e(B);T_{e}\Big]\end{aligned}}{\begin{aligned}\int_0^\infty dE_{\bar{\nu}_e} E_{\bar{\nu}_e}^{2+j} f_{\rm FD}(E_{\bar{\nu}_e},\eta_{\bar{\nu}_e}; T_{\bar{\nu}_e})\end{aligned}}\nonumber,
\end{eqnarray}
where $\eta_{\nu_e} = \mu_{\nu_e}/k_BT$ is the neutrino degeneracy parameter (same definition for $\eta_{\bar{\nu}_e}$), $\sigma(E_n,B)$ is taken from Eq. \ref{sigma_B} and electron chemical potential $\mu_e(B)$ is calculated from Eq. \ref{chemi_po}. $g(E_n,\mu_e;T_e)$ is the Pauli blocking factor for Fermion. $Y_{\rm np}$ and $Y_{\rm pn}$ are coefficients that include the Pauli blocking effects in the phase space of neutrons and protons (See Appendix \ref{appen_a} for those expressions). Since the magnetic moment of nucleons is negligible due to their heavy mass, we ignore the impacts on neutrino scattering processes. One can find their expressions of the cross section and opacity in \cite{Ruffert:1995fs}. For heavy-lepton neutrinos (i.e., $\nu_\tau$ and $\nu_\mu$), only scattering processes are included:
\begin{equation}
\label{k_sca_x}
   \kappa_{{\rm tot},j}(\nu_x) = \kappa_{s,j}(\nu_x n) + \kappa_{s,j}(\nu_x p).
\end{equation}
Finally, the optical depth is given by 
\begin{equation}
\label{nu_sphe}
   \tau_{\nu_i,j}(r)=\int^{\infty}_{r} ds\ \kappa_{{\rm tot}, j}(\nu_i),
\end{equation}
and neutrinosphere by definition, is given by $\tau_{\nu_i,j}(R_\nu) = 2/3$. {Other quantity in leakage scheme such as local emission and absorption rates, effective emission and absorption rates are also modified in terms of an external magnetic field, we put these equations in the Appendix \ref{appen_b}.}
In this work, we use the open-source \texttt{GR1D} code to simulate 1-D SNe explosion \citep{OConnor:2009iuz,OConnor:2014sgn}. We used the LS180 EOS \citep{Lattimer:1991nc} and took a 9.6 M$_\odot$ progenitor model with zero metallicity from \cite{Heger_pri}. The collapse phase employs the parameterization scheme for electron captures following \citet{liebendorfer05} and we switch to the leakage scheme after the core bounce, defined as the moment when entropy per baryon at the edge of the inner core reaches 3\,$k_{\rm B}$. 
{We also note that our 1D simulations do not include magnetic pressure or other magnetic feedback on the  hydrodynamics. For magnetic fields of this strength, such effects could substantially modify the hydrodynamical evolution. In the present work, however, we focus only on the magnetic-field correction to neutrino transport, which should be regarded as an additional effect beyond the hydrodynamics.}

\subsection{Results}
\begin{figure}
\centering
\includegraphics[width=0.5\textwidth,clip=true,trim=0cm 0cm 0cm 0cm]{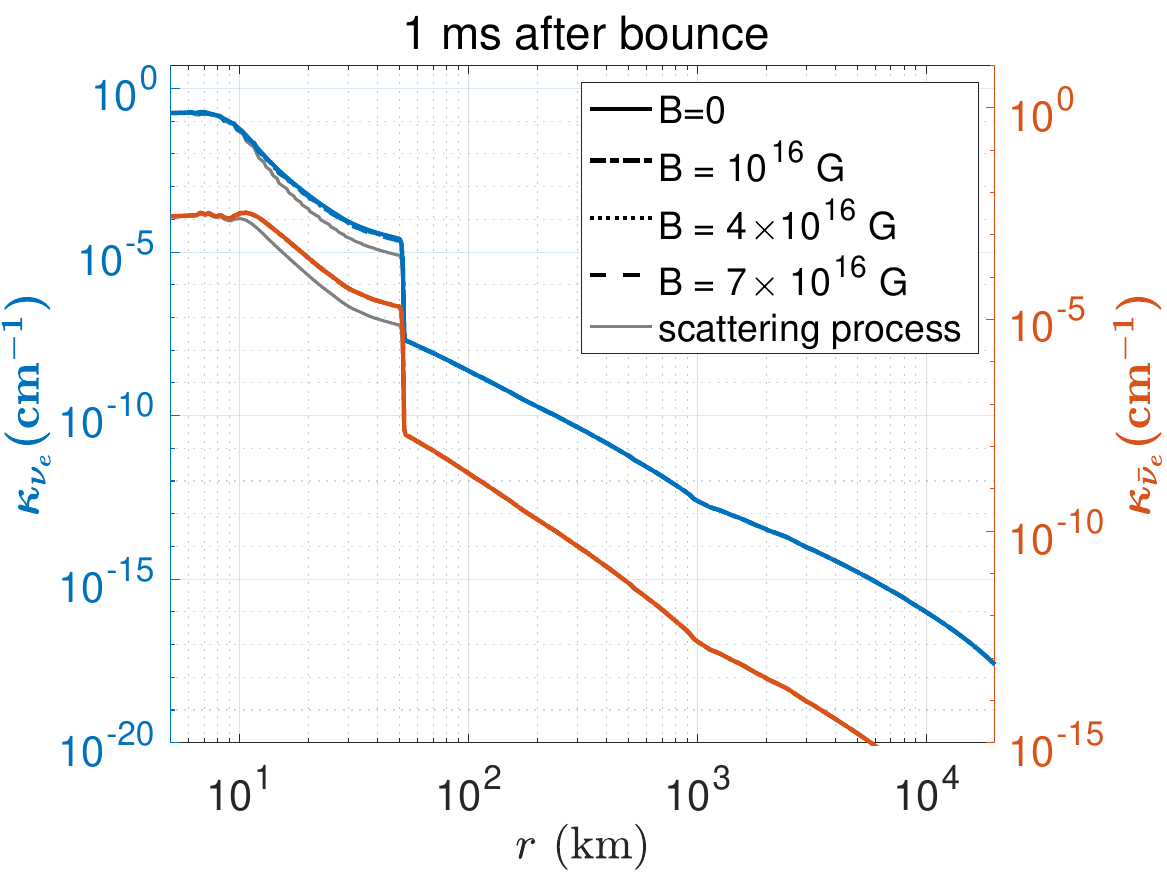}
\includegraphics[width=0.5\textwidth,clip=true,trim=0cm 0cm 0cm 0cm]{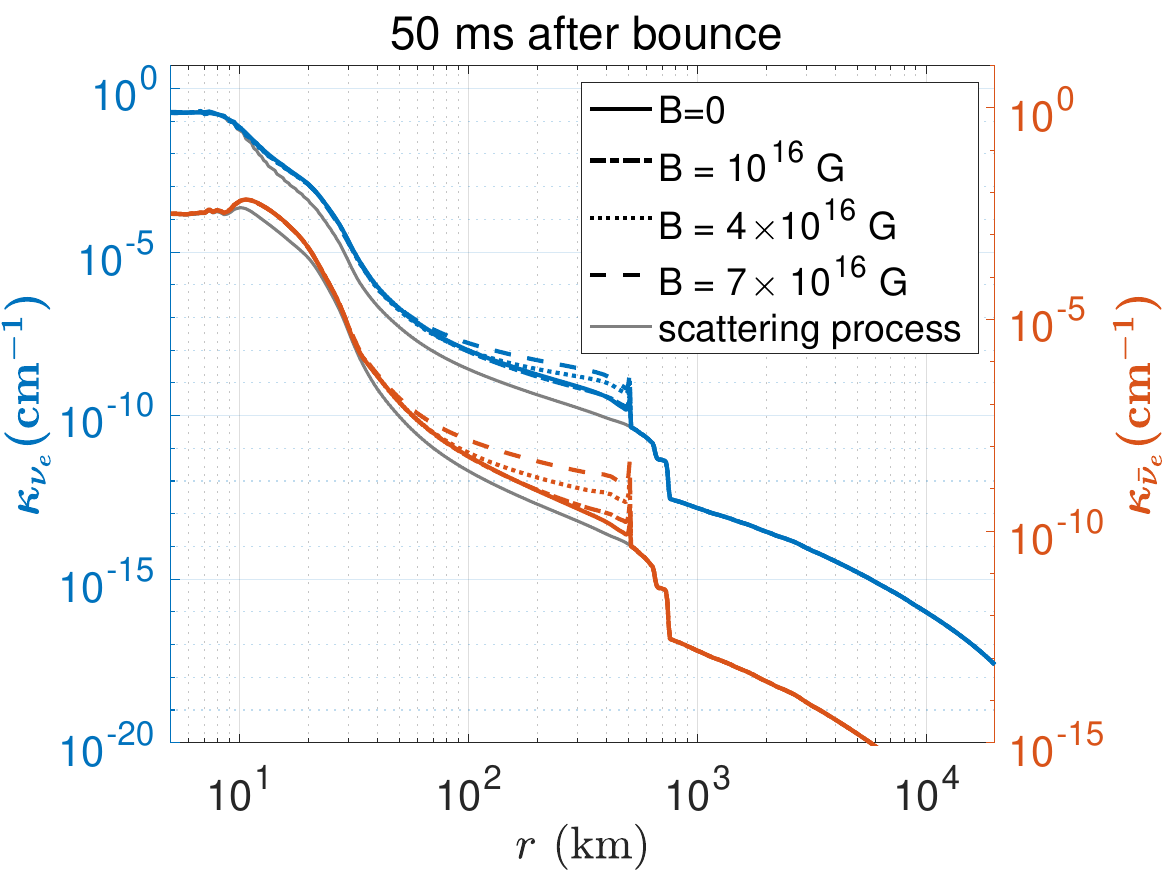}
\includegraphics[width=0.5\textwidth,clip=true,trim=0cm 0cm 0cm 0cm]{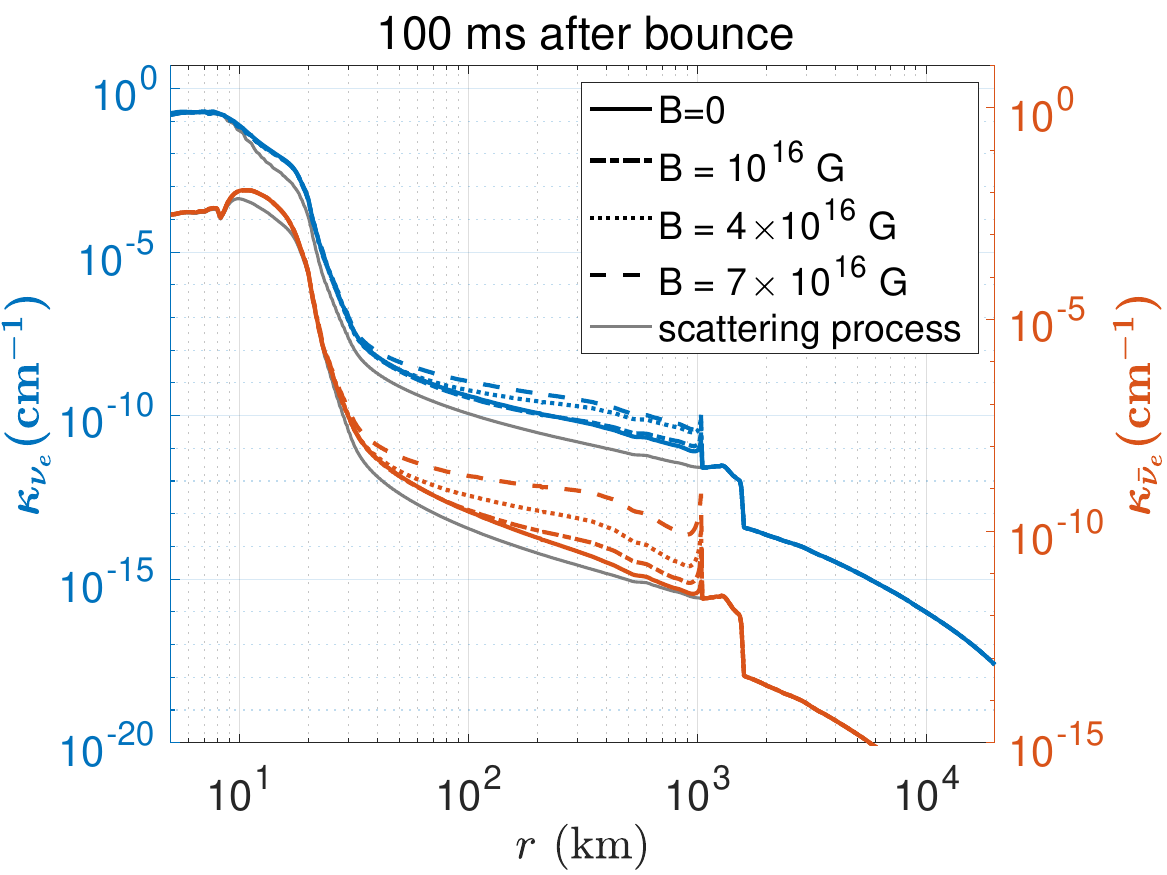}
\caption{\label{k_compare}The opacity of $\nu_e, \bar{\nu}_e$ inside CCSN at different stages, i.e., 1 ms, 50 ms, and 100 ms after core bounce, separately. Blue lines are the opacity for $\nu_e$, and red lines are for $\bar{\nu}_e$. The gray solid lines on all panels represent opacity contributed by scattering processes: $\nu_e + n(p) \to \nu_e + n(p)$ and ${\bar{\nu}}_e + n(p) \to {\bar{\nu}}_e + n(p)$. The line style, i.e., solid, dash-dotted, dotted, and dashed lines, represent the cases of $B = 0$, $B = 10^{16}$ G, $B = 4\times 10^{16}$ G and  $B= 7\times10^{16}$ G, respectively.}
\end{figure}

We first use the extracted density, temperature and $Y_e$ profiles at a certain time step from \texttt{GR1D} code to clarify the magnetic field impacts inside CCSN.
Fig. \ref{k_compare} illustrates the total opacity $\kappa_{\rm tot}$ of neutrinos (blue lines, left $y$-axis) and anti-neutrinos (red lines, right $y$-axis) during different stages (1 ms, 50 ms, and 100 ms after core bounce, respectively) of CCSN. In the inner region of CCSN, neutrino scattering contributes opacity the most, so the modification of absorption processes can be ignored. 

At a very early phase after core bounce, shock propagates in high-density and high-temperature region, so the magnetic field does not make significant contributions, i.e., there is no difference between each line for various magnetic field strengths (upper panel of Fig. \ref{k_compare}). Magnetic field enhances absorption opacity when shock propagates to $ r \geq 30$ km (middle and lower panel of Fig. \ref{k_compare}). This is mainly caused by two effects: 1) Magnetic field quantizes the phase space of $e^\pm$, with extra magnetic energy contribution, which enhances the neutrino-nucleon interaction rates. 
2) The electron chemical potential $\mu_e$ is suppressed since only LLL is occupied, as we discussed in Sec. \ref{B_impact}, leading to a larger value of blocking factor $g(E_n,-\mu_e; T_{e})$ than $g(E_n,\mu_e; T_e)$. These impacts are consistent with those in M1 scheme, as we discussed in \cite{luo25}, i.e., both the cross section deviation and chemical potential modification have more pronounced effect on $\bar{\nu}_e$, leading to a greater enhancement of $\bar\nu_e$ opacity compare to that of $\nu_e$.

At a relatively later stage ($t\sim 100$ ms), the matter cools down and the magnetic field could enhance opacity by about 2-3 orders of magnitude in the region $100$ km $<r<$ $1000$ km inside CCSN (lower panel of Fig. \ref{k_compare}). However, optical depth in such a later phase is mainly determined by the opacity of the inner region ($r<30$ km) due to the sharp decay of opacity with radius. Therefore, the neutrinosphere would eventually become indistinguishable from the field-free scenario (see Fig. \ref{sphere_evol} below).

\begin{figure}
\centering
\includegraphics[width=0.5\textwidth,clip=true,trim=0cm 0cm 0cm 0cm]{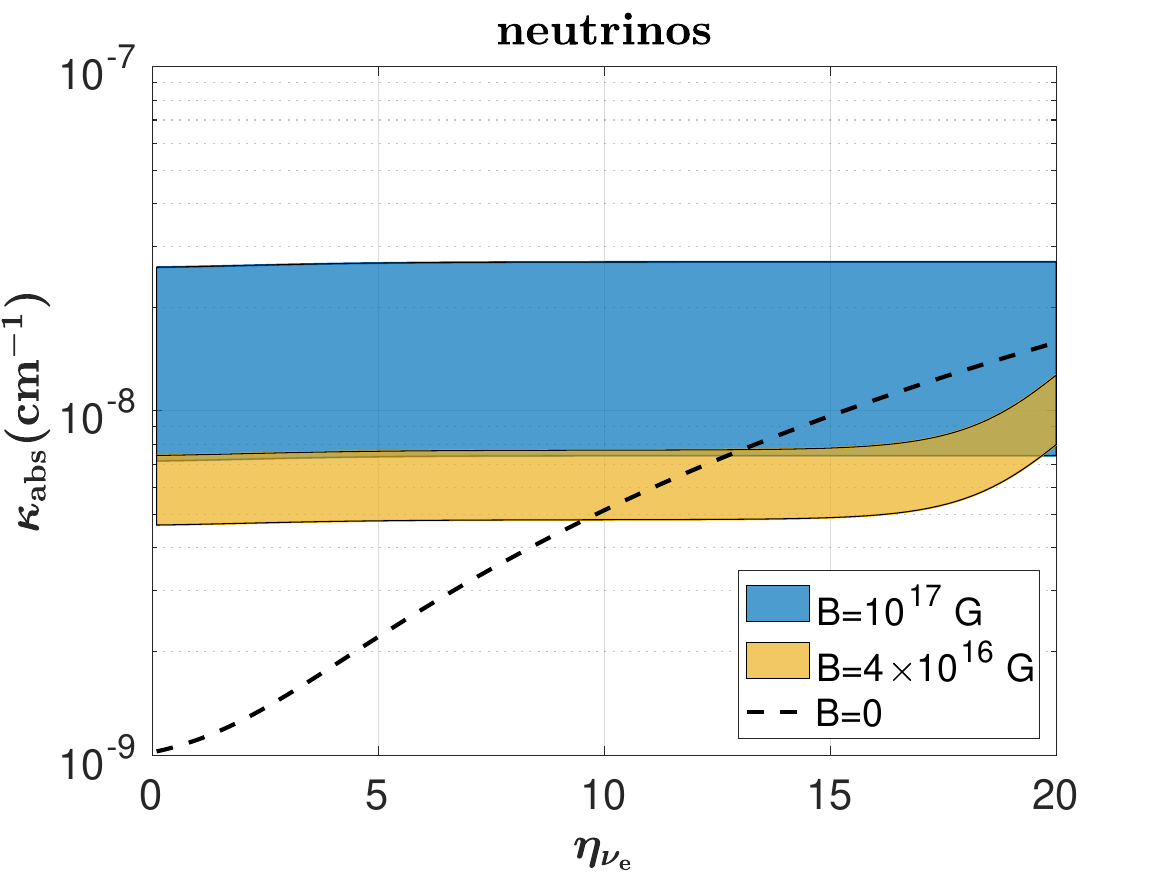}
\caption{\label{kappa_compare}The neutrino absorption opacity $\kappa_{\rm abs}$ under different magnetic field strength. The black solid line is the field-free case, the blue band corresponds to $10^{17}$ G with different values of $\Theta_\nu$ (the lower line of the band is the case that $\Theta_\nu=0$, the upper line of the band is the case that $\Theta_\nu=\pi$), yellow band corresponds to $4\times 10^{16}$ G with the same meaning of the lower and upper lines. In this figure, $T=1$ MeV and $\rho= 10^9\rm \ g\cdot cm^{-3}$. }
\end{figure}

Notice that in a more realistic dipole field case, the $\Theta_\nu$ in Eq. \ref{sigma_B} depends on the neutrino momentum direction and field direction. We apply the different $\Theta_\nu$ to calculate the angle dependence of $\kappa_{abs}(\nu_e)$ as a function of $\eta_{\nu_e}$. The results are shown in Fig. \ref{kappa_compare}. Here, the width of blue and yellow bands are caused by changing the value of $\Theta_\nu$ from 0 to $\pi$. The temperature and density are set to be $T=1$ MeV and $\rho= 10^9\rm \ g\cdot cm^{-3}$. The absorption opacity can differ by a factor of 2 by changing the value of $\Theta_\nu$ for $B=10^{17}$ G. However, this is the upper limit of the magnetic field strength. The yellow band shows less than 20\% difference for the case of a $4\times10^{16}$ G magnetic field, indicating that neglecting the angular effect can be a reasonable approximation, so we set $\Theta_\nu=0$ through this work.

\begin{figure}
\centering
\includegraphics[width=0.5\textwidth,clip=true,trim=0cm 0cm 0cm 0cm]{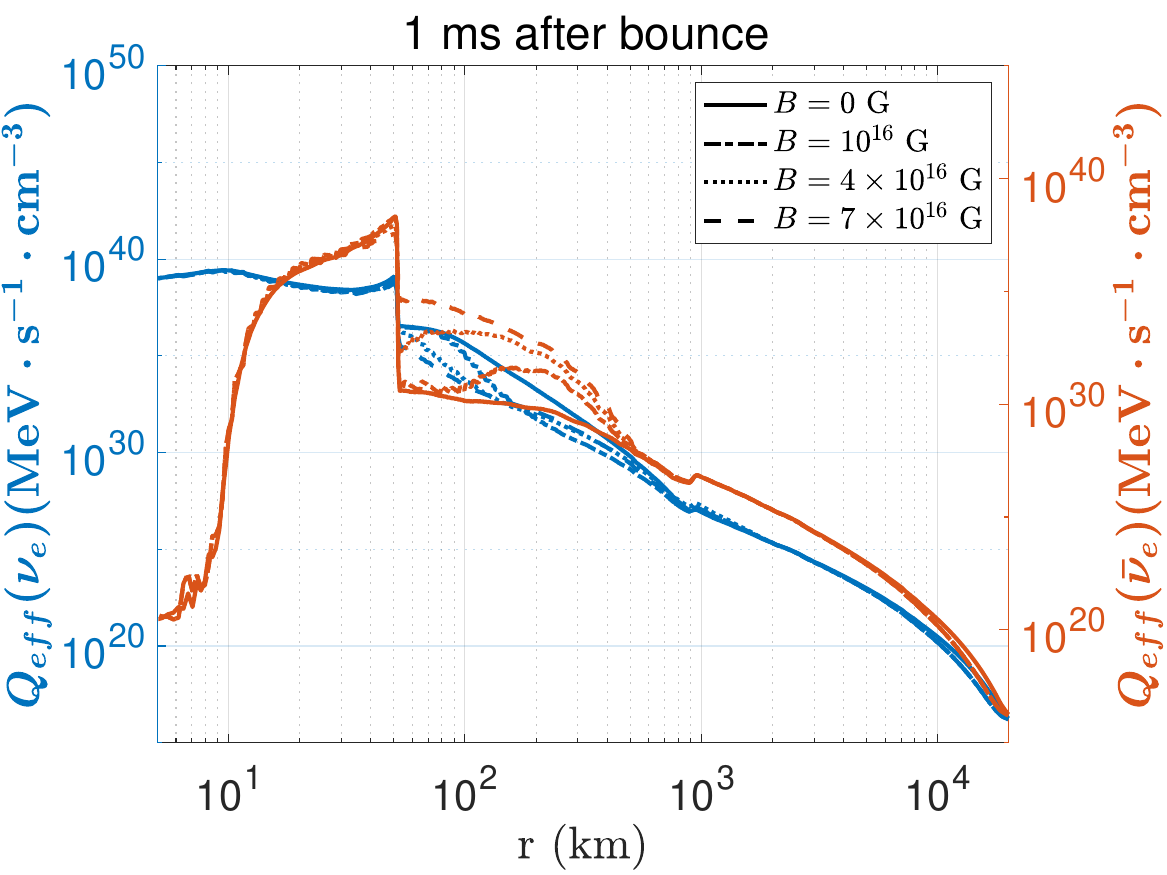}
\includegraphics[width=0.5\textwidth,clip=true,trim=0cm 0cm 0cm 0cm]{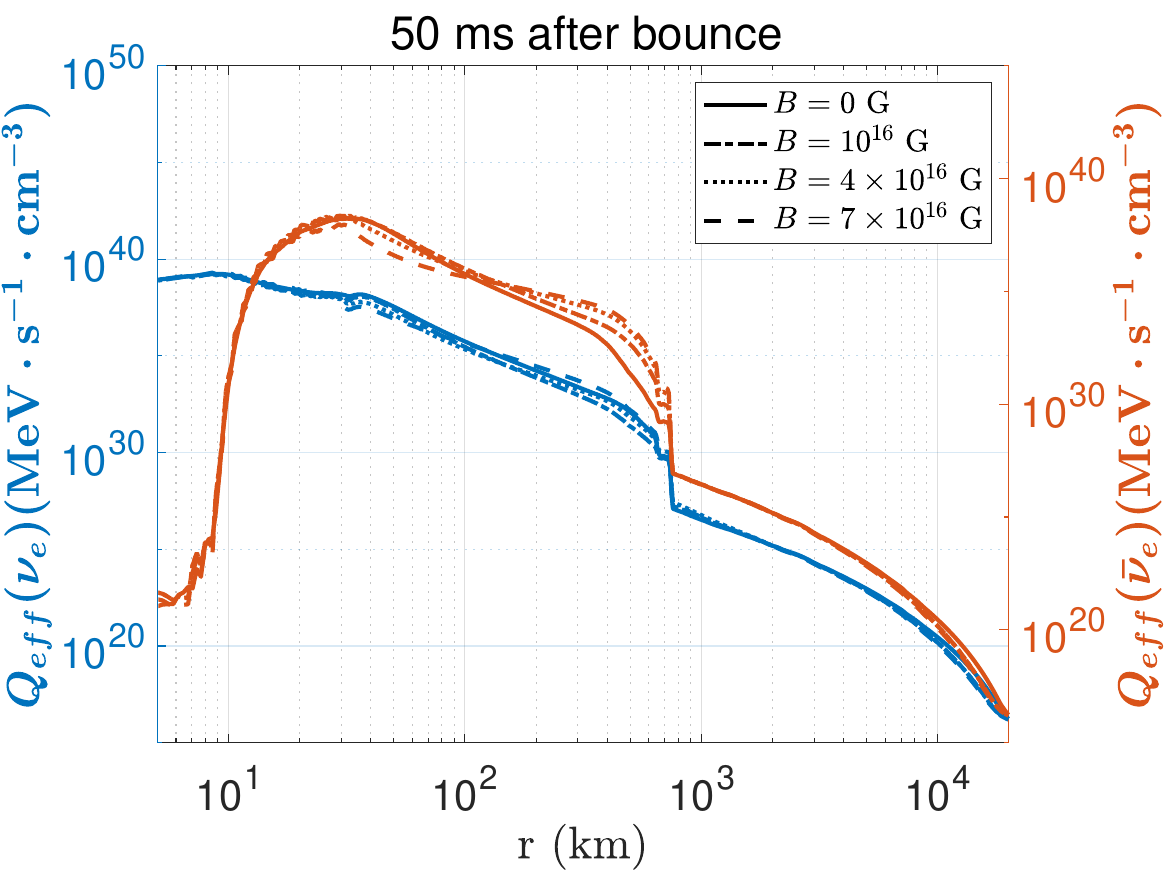}
\includegraphics[width=0.5\textwidth,clip=true,trim=0cm 0cm 0cm 0cm]{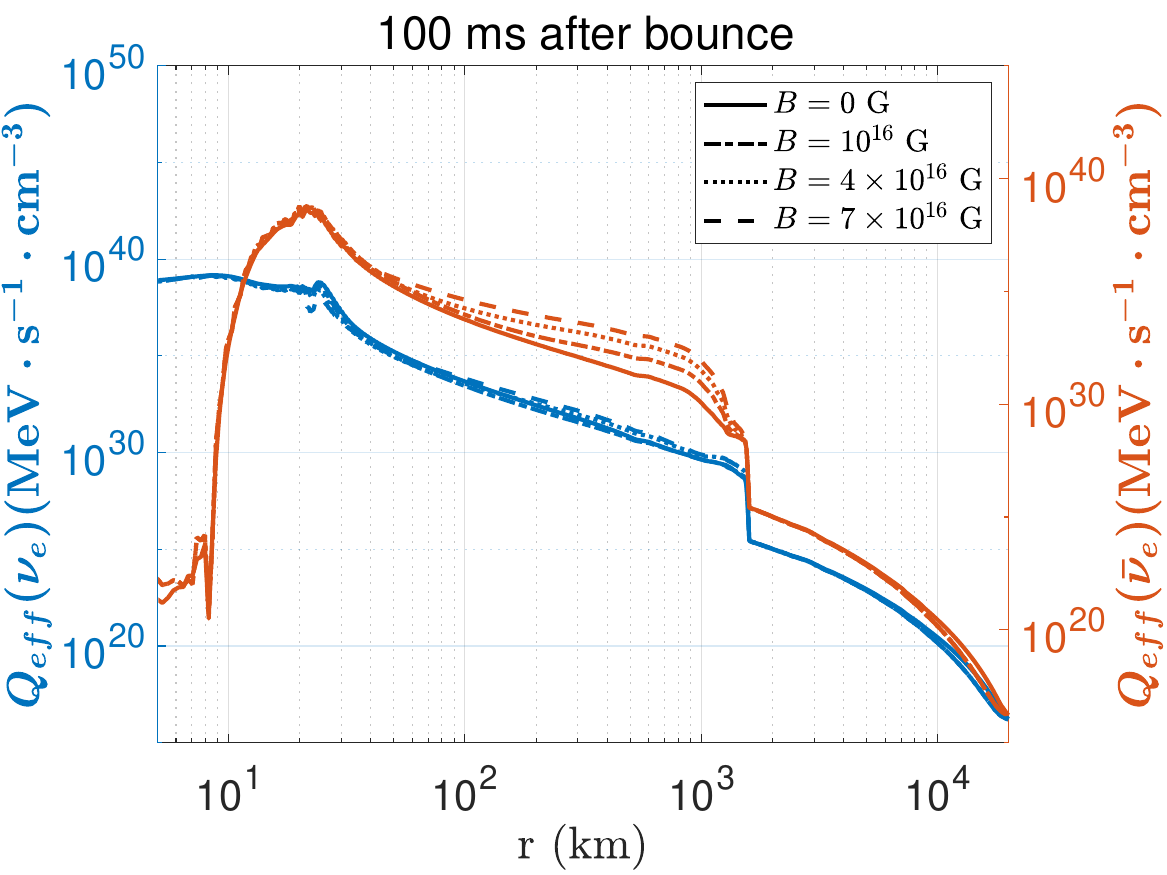}
\caption{\label{eff_compare}The effective energy emission rate of $\nu_e$ and $\bar{\nu}_e$ inside the CCSN after core bounce. The line-color convention follows Fig. \ref{k_compare}.}
\end{figure}

The effective neutrino emission rates Eq. \ref{RQ_eff} describes whether the neutrinos are diffused or emitted inside a CCSN: for a high-density regime, diffusion rates are high, so it governs the main neutrino number and energy loss. In contrast, the main process becomes the local emission rate if neutrinos travel to a low-density region. 
{We show the effective emission rates} in Fig. \ref{eff_compare} for different stages of a CCSN (from up to bottom panel: 1 ms, 50 ms and 100 ms after the core bounce, respectively). Again, we compare the impacts of magnetic field under the extracted density, temperature and $Y_e$ conditions in this figure. About 50 ms after core bounce (middle panel of Fig. \ref{eff_compare}), a magnetic field suppresses the effective neutrino emission rates in the range of $30\ {\rm km}<r< 100 \ {\rm km}$, which is due to the enhancement of opacity we have shown in the middle panel of Fig. \ref{k_compare}. Also, such a result is consistent with the reduction of the diffusion rates in this region (cf. Appendix \ref{appen_b}). At 100 ms after core bounce (lower panel of Fig. \ref{eff_compare}), shock has already propagated to the outer region, so the matter cools down and has a lower density. Diffusion is not essential in such an environment, while the local emission rates are enhanced more drastically (cf. Appendix \ref{appen_b}), leading to stronger effective emission rates.

We further modified the neutrino transport in \texttt{GR1D} based on Eq. \ref{k_abs1}, Eq. \ref{k_abs2} and equations in Appendix \ref{appen_b} to perform 1-D simulation with magnetic field. We investigate impacts from a constant magnetic field in a region $r<r_{\rm cut}=100$ km in our CCSN simulation.
The enhanced opacity enlarges the radius of neutrinosphere, which is shown in Fig. \ref{sphere_evol} where time evolution of neutrinosphere for $\nu_e$  and $\bar{\nu}_e$ are shown in the upper and lower panel, respectively (our calculation uses the opacity for neutrino-energy transport $\kappa_{{\rm tot},1}(\nu_i)$ to determine the neutrinosphere). The radius of neutrinosphere is indistinguishable under different magnetic field cases at $t<10$ ms after core bounce. After a few dozens ms, when shock propagates outside the high-density and high-temperature region, $\kappa_{\rm tot}$ increases and one can see the enlargement of neutrinosphere as expected. Usually, without a magnetic field, the radius of the neutrinosphere is larger than that of anti-neutrinos. This is because the proton abundance reduces after the shock passage, leading to a smaller value of absorption opacity $\kappa_{a}$. However, for a magnetic field as strong as $7\times 10^{16}$ G (i.e., blue lines), $R_{\bar{\nu}_e}$becomes comparable with $R_{\nu_e}$, this is consistent with the fact that magnetic field enlarges $\kappa_{a}(\bar{\nu}_e)$ more than $\kappa_{a}({\nu}_e)$ as we discussed above.
\begin{figure}
\centering
\includegraphics[width=0.5\textwidth,clip=true,trim=0cm 0cm 0cm 0cm]{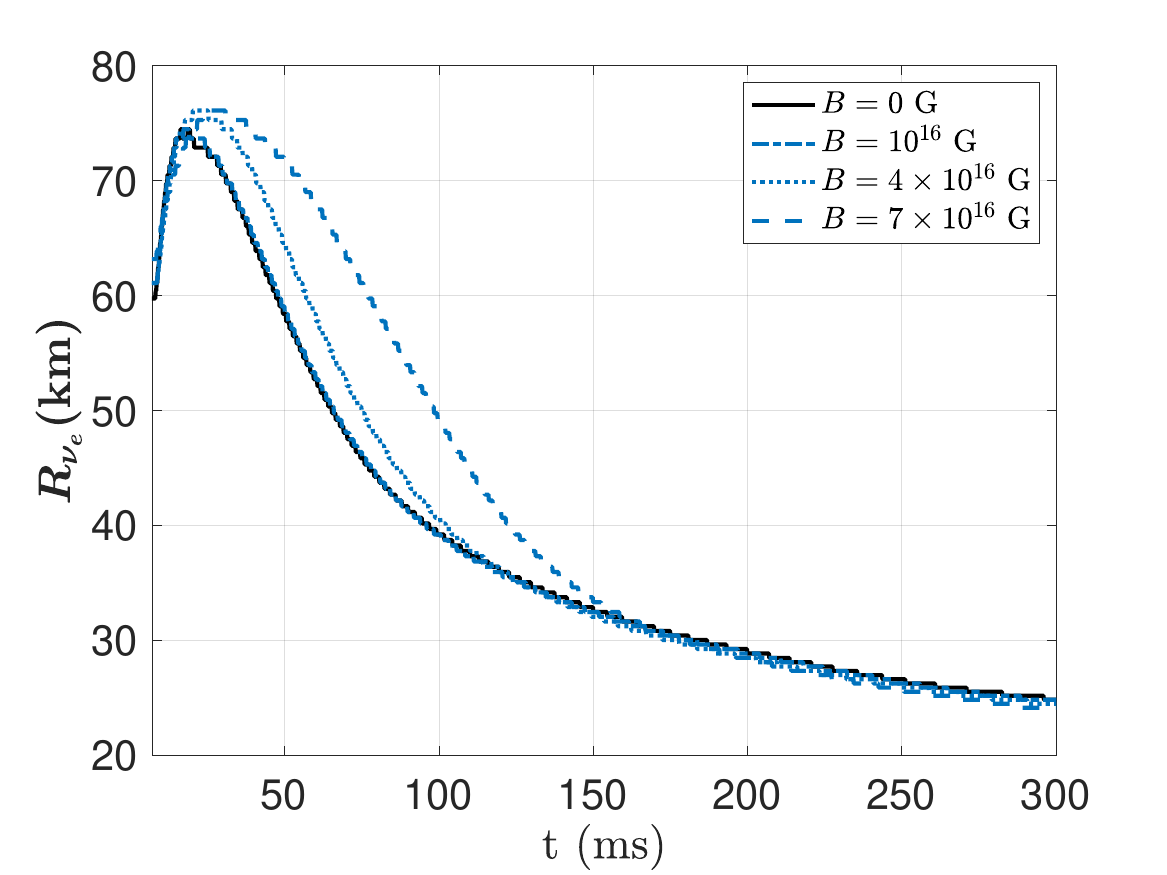}
\includegraphics[width=0.5\textwidth,clip=true,trim=0cm 0cm 0cm 0cm]{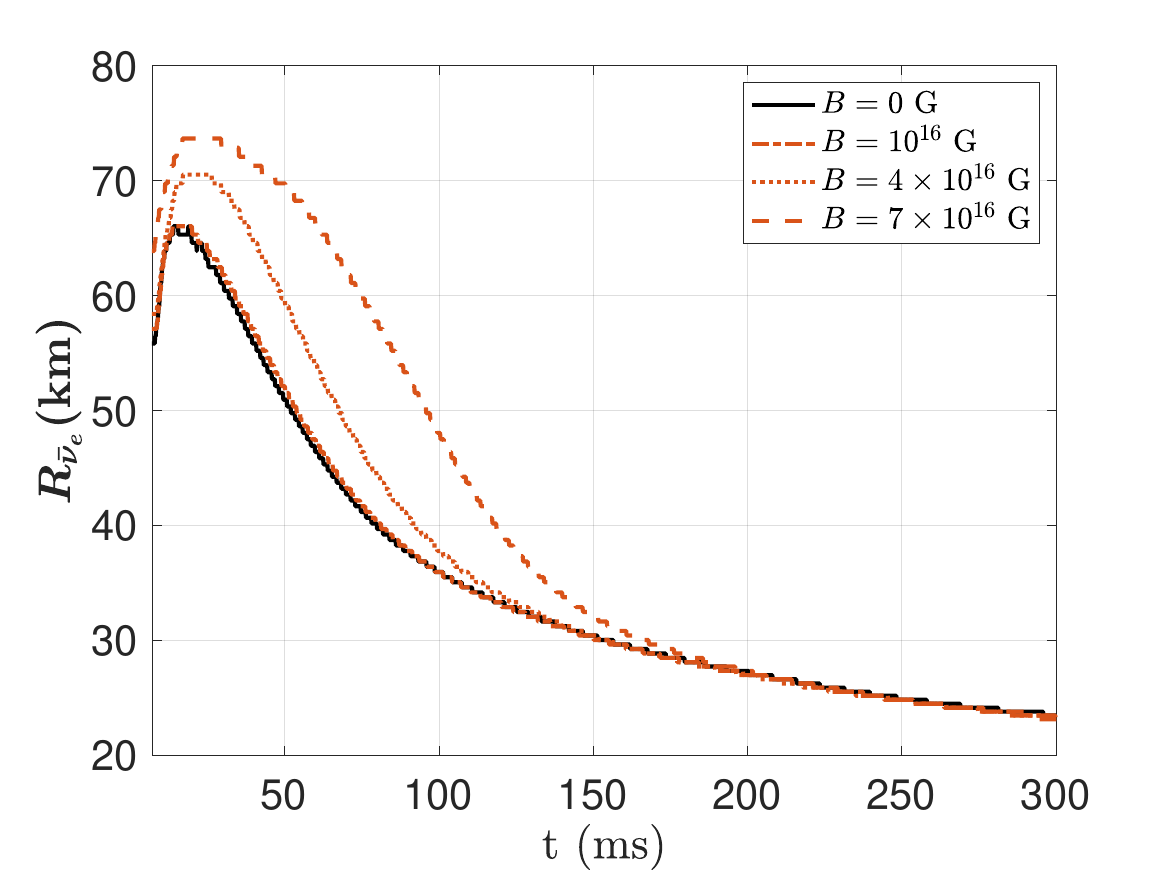}
\caption{\label{sphere_evol}Upper panel: The neutrinosphere evolution ($\nu_e$) after core bounce. Lower panel: The neutrinosphere evolution of $\bar{\nu}_e$ after core bounce. Black solid lines corresponding to the field free case; red, green, and blue lines corresponding to $B=10^{16}$ G,  $B=4\times 10^{16}$ G and  $B=7\times 10^{16}$ G, respectively.}
\end{figure}

\begin{figure}
\centering
\includegraphics[width=0.5\textwidth,clip=true,trim=0cm 0cm 0cm 0cm]{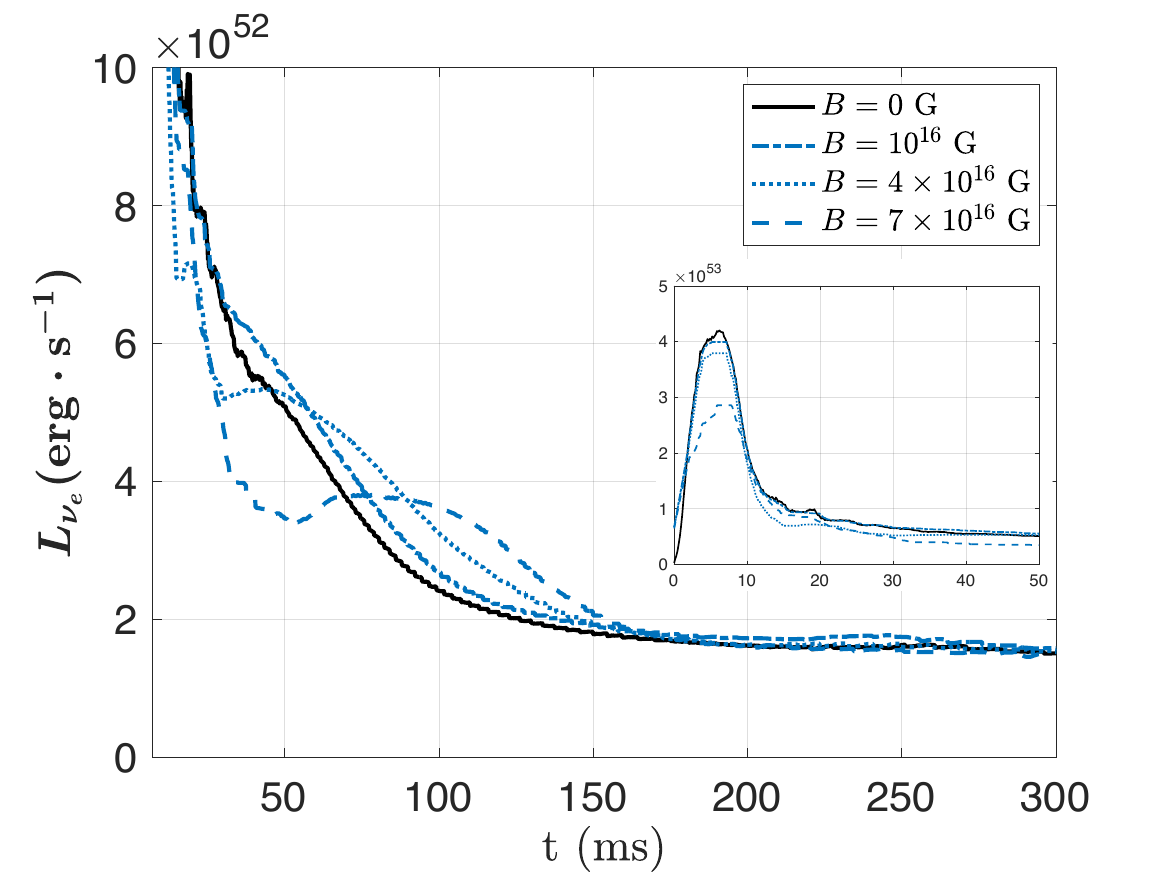}
\includegraphics[width=0.5\textwidth,clip=true,trim=0cm 0cm 0cm 0cm]{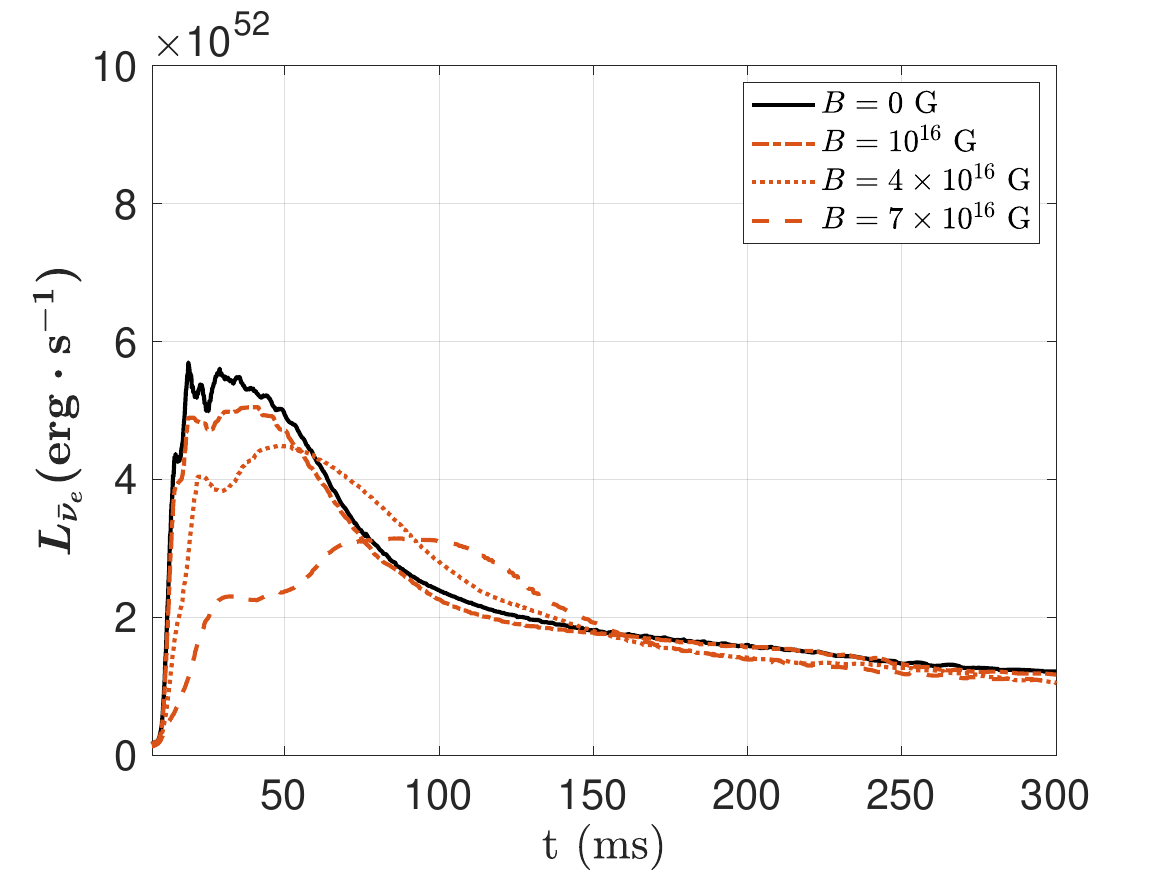}
\caption{\label{lumi}Upper panel: The neutrino luminosity ($L_{\nu_e}$) after core bounce. Lower panel: The anti-neutrino luminosity ($L_{\bar{\nu}_e}$) after core bounce. The line-color convention follows Fig. \ref{sphere_evol}.}
\end{figure}

The effective emission rates further determine neutrino luminosity, as shown in Fig. \ref{lumi}. After core bounce, $\nu_e$ is created by the electron capture on protons in the heated matter. A luminous flash of neutronization $\nu_e$ is radiated when shock travels to the low-density region outside the neutrinosphere. In field-free case, $L_{\nu_e}$ reaches the peak value near $4\times 10^{53} \rm \  erg\cdot s^{-1}$ at $\sim 7$ ms (black solid line of the upper panel, also see the insert figure on the upper panel). The peak of $L_{\bar{\nu}_e}$ came shortly later than neutrinos (black solid line of the lower panel), with the value about $6\times 10^{52} \rm \ erg\cdot s^{-1}$ at $\sim 30$ ms; this is due to $\bar{\nu}_e$ production by pair process in shock-heated matter at a later time. 

The situation changed after introducing the magnetic field: the peak luminosity decreases (red, green, and blue lines in each panel of Fig. \ref{lumi}). Especially, $L_{\bar{\nu}_e}$ peak value is more sensitive to the magnetic field. This is reasonable since in each panel of Fig. \ref{eff_compare}, effective emission rates of $\bar{\nu}_e$ show more significant change than that of ${\nu}_e$.
More interestingly, since the magnetic field mainly enlarges the absorption rate of neutrinos, i.e. neutrinos spend more time trapped inside the neutrinosphere, therefore, after reaching the peak luminosity, both $L_{{\nu}_e}$ and $L_{\bar{\nu}_e}$ decay time scales become longer. 

Furthermore, the integrated energy of $\nu_e$ until 300 ms is reduced by about 10\% ($1.13\times 10^{52}$ erg for field-free case, $1.03\times 10^{52}$ erg for a $7\times 10^{16}$ G magnetic field). While for $\bar{\nu}_e$, more trapping time leads to  about 20\% reduction of integrated energy ($6.92\times 10^{51}$ erg and $5.52 \times 10^{51}$ erg for field-free and $7\times 10^{16}$ G cases, respectively).

\begin{figure}
\includegraphics[width=0.5\textwidth,clip=true,trim=0cm 0cm 0cm 0cm]{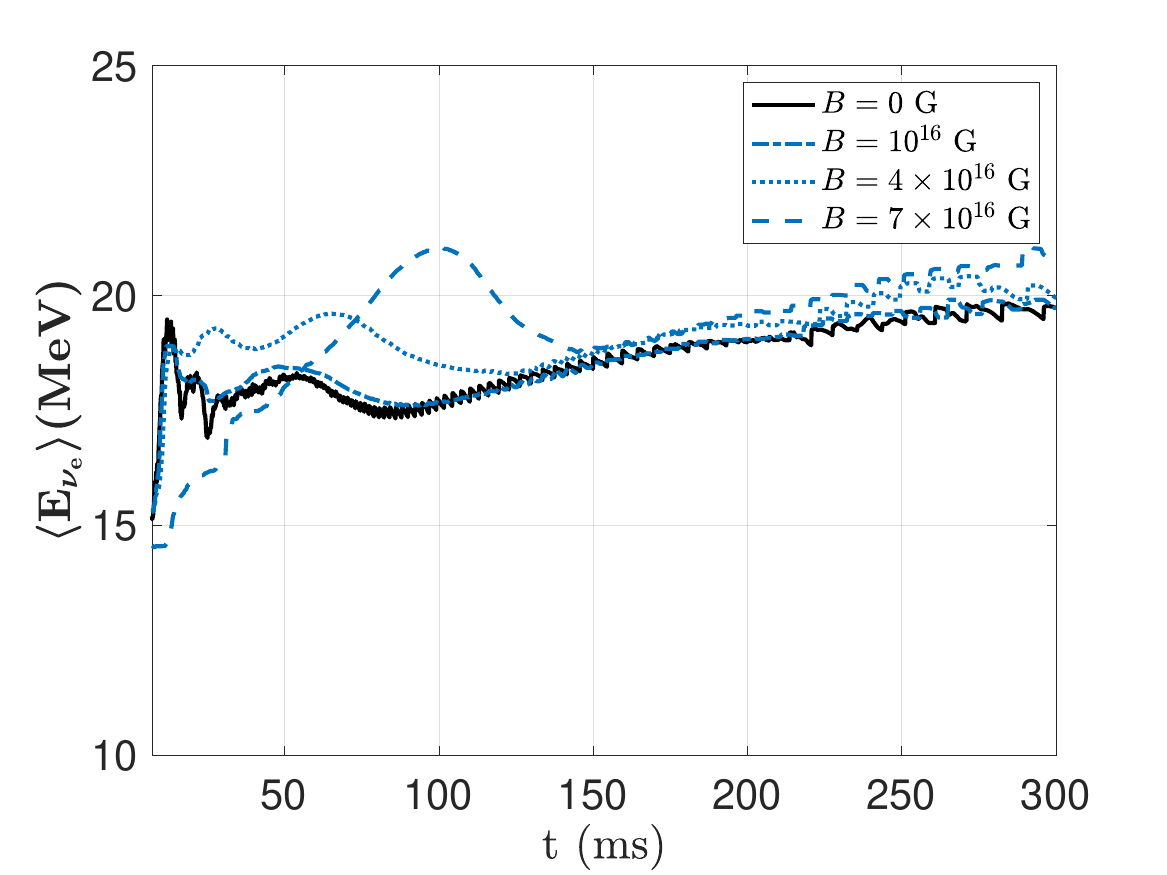}
\includegraphics[width=0.5\textwidth,clip=true,trim=0cm 0cm 0cm 0cm]{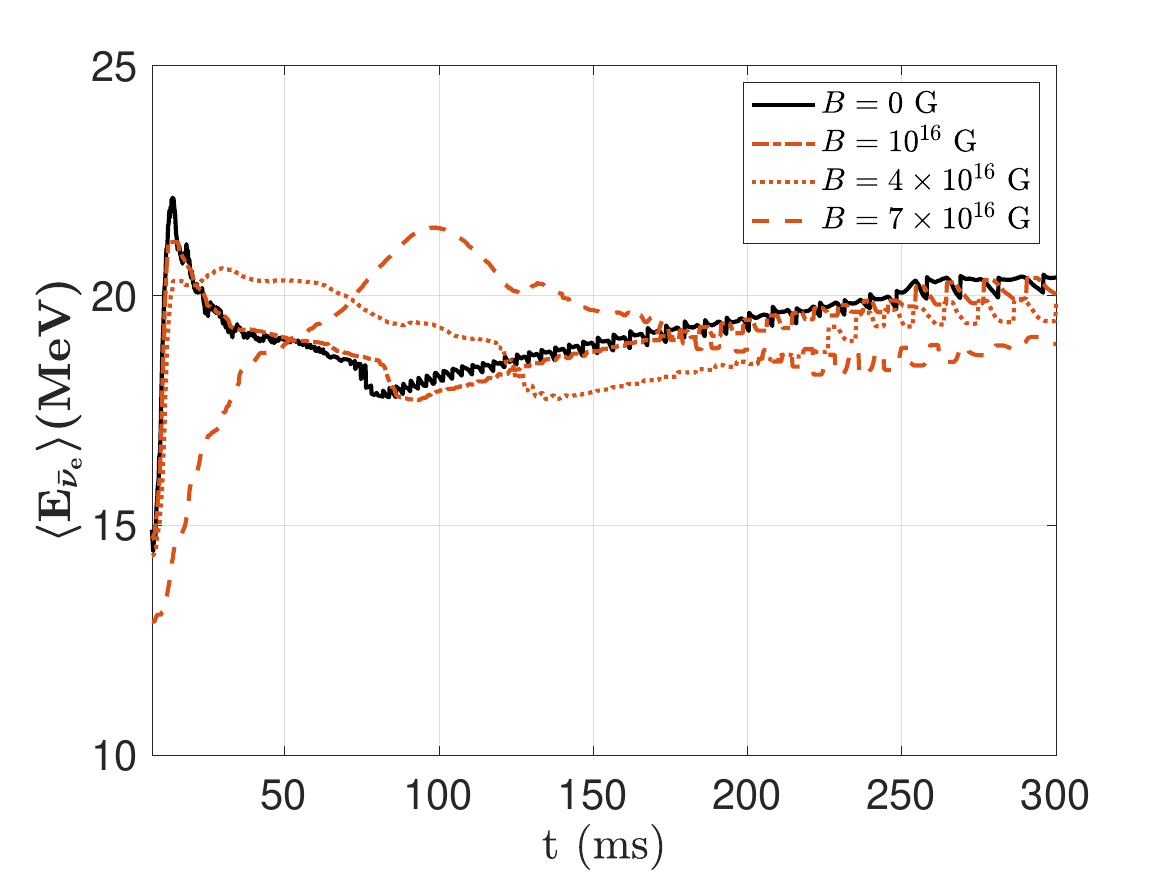}
\caption{\label{engy}Upper panel: The neutrino mean energy ($\langle E_{\nu_e}\rangle$) as a function of time after core bounce. Lower panel: The anti-neutrino mean energy ($\langle E_{\bar{\nu}_e}\rangle$) as a function of time after core bounce. The line-color convention follows Fig. \ref{sphere_evol}.}
\end{figure}

Fig. \ref{engy} shows the time evolution of neutrino mean energies $\langle E_{\nu_e}\rangle$ (upper panel) and $\langle E_{\bar{\nu}_e}\rangle$ (lower panel) at the corresponding neutrinospheres. Because the employed leakage scheme assumes the neutrino energy spectra to be Fermi-Dirac, the neutrino mean energies are tightly correlated with the local matter temperatures. Thus the dependence of their time evolution on the magnetic field strength reflects the corresponding dependence of the neutrinosphere evolution (see Fig.~\ref{sphere_evol}). With a stronger magnetic field, the neutrino mean energies are smaller shortly after bounce and reach their peak values at later times. Note that this gray neutrino leakage scheme cannot accurately predict their energy spectra and mean values. Nevertheless, the general dependence on the magnetic field strength found here should still hold in simulations with more accurate multi-energy group neutrino transport. 

\begin{figure}
\centering
\includegraphics[width=0.5\textwidth,clip=true,trim=0cm 0cm 0cm 0cm]{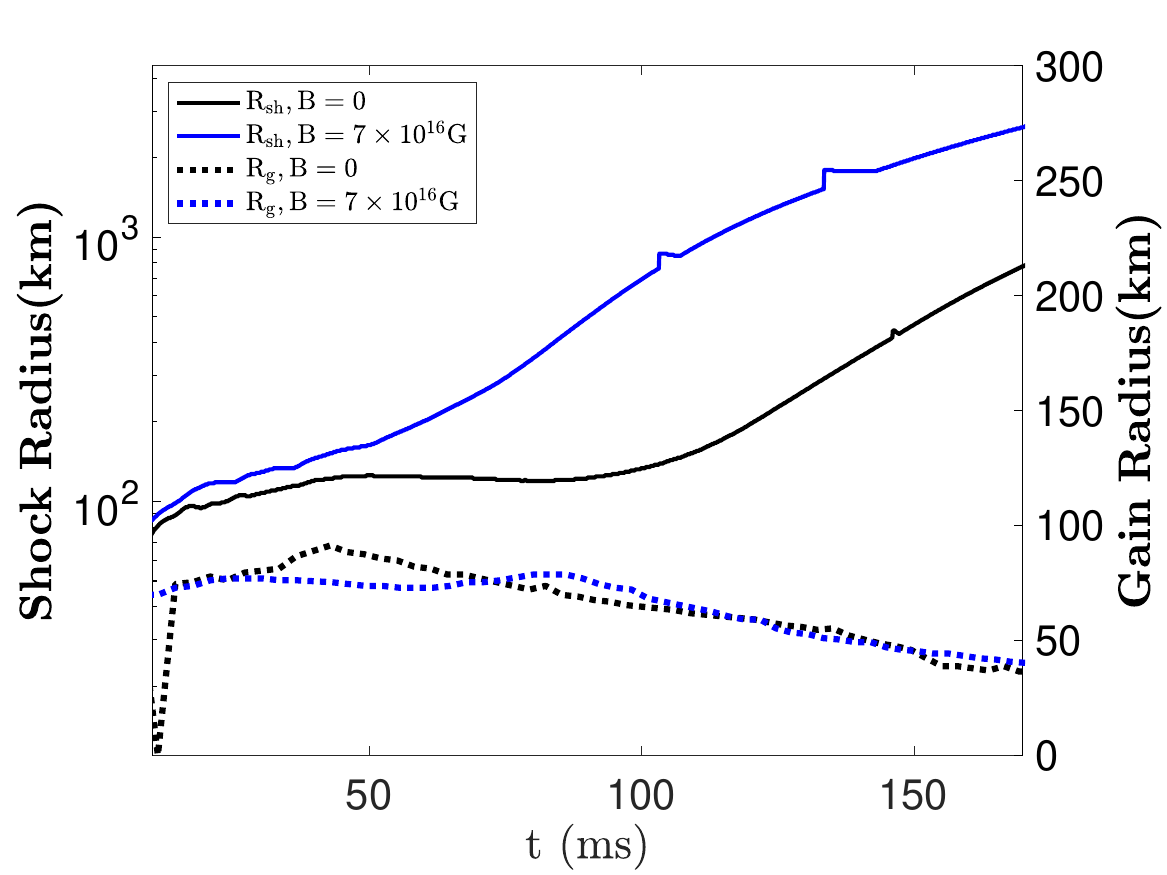}
\caption{\label{shock_rg} The comparison of shock radius $R_{\rm sh}$ (solid lines) and gain radius $R_{\rm g}$ (dotted lines) under field-free case (black lines) and $7\times10^{16}$ G (blue lines). The gain radius separates the net neutrino cooling (below $R_{\rm g}$) and heating (above $R_{\rm g}$) regions.}
\end{figure}

We further inspect the impacts of the modified neutrino transport by magnetic field on the neutrino heating and explosion dynamics. Fig.~\ref{shock_rg} compares the evolution of the shock radius ($R_{\rm sh}$, left $y$-axis) and gain radius ($R_{\rm g}$, right $y$-axis) between the field-free and strongest field cases. The gain radius separates the net neutrino cooling and heating regions and its location and time evolution are crucial for the explodability of CCSNe. Before $\sim75$\,ms postbounce, $R_{\rm g}$ is smaller by at most $\sim20$\,km in the case of a strong magnetic field than that in the field-free case. Accordingly, the runaway shock expansion takes place earlier ($\sim60$\, ms postbounce) in the case of a strong magnetic field than in the field-free case ($\sim120$\, ms postbounce). During $\sim75-100$\,ms, $R_{\rm g}$ in the case of a strong magnetic field becomes larger than in the field-free case due to the earlier runaway shock expansion. After $\sim100$\,ms postbounce, $R_{\rm g}$ is almost identical for the two cases because it has receded to below $\sim50$\,km where magnetic field does not affect the neutrino opacities (cf.~Fig.~\ref{k_compare}) and transport (cf.~Fig.~\ref{eff_compare}). 

\section{Conclusion}\label{con_dis}
In this study, we derived the neutrino leakage scheme formula under the influence of strong magnetic fields and employed the \texttt{GR1D} code to perform a 1-D CCSN simulation of a 9.6 M$_{\odot}$ zero-metallicity progenitor. The presence of a magnetic field modifies the neutrino absorption cross sections and diminishes the $e^\pm$ chemical potential, as these particles are restricted to the lowest Landau level. We studied the effects of a constant magnetic field within a specific radius configuration, assuming field strengths ranging from $10^{16}$ G to $7 \times 10^{16}$ G during the post-bounce phase.

The magnetic field enhances the absorption rates of neutrinos and anti-neutrinos, enlarging their opacities (Fig. \ref{k_compare}). This is consistent with results from the M1 scheme, where both cross-section deviations and chemical potential modifications have more pronounced effects on $\bar{\nu}_e$, leading to a greater enhancement of $\bar{\nu}_e$ opacity compared to $\nu_e$. The larger opacity increases the trapping timescale of neutrinos inside the neutrinosphere, resulting in longer decay timescales for neutrino luminosities and smaller peak luminosities (Fig. \ref{lumi}). Accordingly, with stronger magnetic fields, neutrinos exhibit lower mean energies shortly after bounce and reach their peak values at later times (Fig. \ref{engy}). 

These results demonstrate that the leakage scheme provides an efficient and practical framework for studying the qualitative impacts of magnetic fields on neutrino transport in 1D CCSNe simulations.

\begin{acknowledgements}
Y. Luo is supported by the China Postdoctoral Science Foundation (Grant. No. 2025T180924), the National Natural Science Foundation of China (NSFC, No. 12335009) and the Boya Fellowship of Peking University. S. Zha is supported by the National Natural Science Foundation of China (NSFC, Nos. 12288102, 12393811, 12473031), the International Centre of Supernovae, Yunnan Key Laboratory (No. 202505AV340004), the Yunnan Fundamental Research Project (Nos. 202401BC070007, 202501AS070078) and the Yunnan Revitalization Talent Support Program--Young Talent project. T. Kajino is supported in part by the National Key R\&D Program of China (2022YFA1602401), the National Natural Science Foundation of China (No. 123352, 12393009), and Grants-in-Aid for Scientific Research of Japan Society for the Promotion of Science (20K03958).
\end{acknowledgements}
\appendix
\section{Fermion blocking effects of nucleon}\label{appen_a}
In Eq. \ref{k_abs1} and Eq. \ref{k_abs2}, $Y_{\rm np}$ and $Y_{\rm pn}$ are given by \citep{Ruffert:1995fs,Rosswog:2003rv}:
\begin{equation}
   Y_{\rm np} = \dfrac{2Y_e -1 }{ \exp{(\eta_p-\eta_n)}-1};
   Y_{ \rm pn}  = \exp{(\eta_p-\eta_n)} \cdot Y_{\rm np},
\end{equation}
where $\eta_{n(p)} = \mu_{n(p)}/k_BT$ is the degeneracy parameter of protons and neutrons. The neutrino degeneracy parameters $\eta_\nu$ are determined from local optical depth $\tau_{\nu_i}(r)$:
\begin{equation}
\l\{ 
\begin{aligned}
   &\eta_{\nu_x}(r) = 0\ \ \ \ \ \ \\
   &\eta_{\nu_e}(r) = \eta^{eq}_{\nu_e} \cdot \big[1-\exp{(-\tau_{\nu_e}(r))}\big] \ \ \ \ \ \ \ \\
   &\eta_{\bar{\nu}_e}(r) = - \eta^{eq}_{\bar{\nu}_e} \cdot \big[1-\exp{(-\tau_{\bar{\nu}_e}(r))}\big].
 \end{aligned} \r.
\end{equation}
Here, $\eta^{eq}$ is the degeneracy parameter in the equilibrium state following the relation:
\begin{equation}
   \eta^{eq}_{\nu_e} = -\eta^{eq}_{\bar{\nu}_e} = \eta_e(B) +\eta_p - \eta_n -Q/k_BT, 
\end{equation}
where $Q = m_n-m_p$. 
\section{Neutrino diffusion and emission rates}\label{appen_b}
Under a magnetic field, local emission rates from $e^\pm$ capture are given by
\begin{eqnarray}
\label{emit_eq}
   &&R^B_{e,j}(e^- p) = A\rho Y_{\rm np}\times\\
   &&\int_0^\infty dE_{\nu_e}  \sigma(E_n,B)E_{\nu_e}^{2+j} g(E_{\nu_e}, \eta_{\nu_e}; T_{\nu_e})f_{\rm FD}\Big[E_n, \mu_e(B);T_e\Big],\nonumber\\
   \nonumber\\
   &&R^B_{e,j}(e^+ n) = A\rho Y_{\rm pn} \times\\
   &&\int_0^\infty dE_{\bar{\nu}_e}\sigma(E_n,B)E_{\bar{\nu}_e}^{2+j} g(E_{\bar{\nu}_e}, \eta_{\bar{\nu}_e}; T_{\bar{\nu}_e}) f_{\rm FD}\Big[E_n, -\mu_e(B);T_{e}\Big], \nonumber
\end{eqnarray}
where $j=0$ is number emission rate, and $j=1$ is energy emission rate. Pair creation ($e^- +e^+ \to \nu_e +\bar\nu_e$) and Bremsstrahlung effect are also included in \texttt{GR1D} code, we keep these two types of $\nu_e (\bar\nu_e)$ emission rates as field-free scenarios. Similar to \cite{Rosswog:2003rv,Ruffert:1995fs}, this work uses the effective emission rates in leakage scheme:
\begin{equation}
\label{RQ_eff}
   R_{\rm eff} =\dfrac{ R_{\rm emit} }{ 1+R_{\rm emit}/R_{\rm diff}},\ Q_{\rm eff} =\dfrac{ Q_{\rm emit} }{ 1+Q_{\rm emit}/Q_{\rm diff}},
\end{equation}
where $R_{\rm emit}$ and $Q_{\rm emit}$ are the local neutrino number and energy emission rates, respectively: 
\begin{eqnarray}
   R_{\rm emit} = R^B_{e,0}+R_{\rm Pair}+R_{\rm Brem}, \nonumber\\
   Q_{\rm emit} = R^B_{e,1}+Q_{\rm Pair}+Q_{\rm Brem}.
\end{eqnarray}

The local diffusion rates $R_{\rm diff}$ and $Q_{\rm diff}$ are calculated following the approximation made in \cite{Rosswog:2003rv}:
\begin{eqnarray}
&& R_{\rm diff} = \dfrac{4\pi c g_{\nu_i}}{ (hc)^3}\dfrac{\zeta_{\nu_i}}{ 3\chi^2_{\nu_i}}TF_0(\eta_{\nu_i}),\\
    \nonumber\\
&& Q_{\rm diff} = \dfrac{4\pi c g_{\nu_i}}{ (hc)^3}\dfrac{\zeta_{\nu_i}}{ 3\chi^2_{\nu_i}}T^2F_1(\eta_{\nu_i}),
\end{eqnarray}
for neutrino species $\nu_i$, where $\chi_{\nu_i}=\int \zeta_{\nu_i} dr$. The strong magnetic field changes the neutrino absorption rate on nucleons, i.e., value of $\zeta_{\nu_e}({\nu}_e n)$ and $\zeta_{\bar{\nu}_e}(\bar{\nu}_e p)$ are modified. By definition, these two quantities are the local absorption rate over $\langle E^2\rangle$ when calculating the diffusion rate:
\begin{eqnarray}\label{diffu_rate}
   &&\zeta_{\nu_e}({\nu}_e n) = A\rho Y_{\rm np} \times\\
   &&\dfrac{\int_0^\infty \sigma(E_n, B)E_{\nu_e}^{2} f_{\rm FD}(E_{\nu_e}, \eta_{\nu_e}; T_{\nu_e}) g\Big[E_n, \mu_e(B);T_e\Big] }{ \int_0^\infty E_\nu^{4} f_{\rm FD}(E_{\nu_e}, \eta_{\nu_e}; T_{\nu_e})},\nonumber\\
   &&\zeta_{\bar{\nu}_e}(\bar{\nu}_e p)= A\rho Y_{\rm np} \times\\
   &&\dfrac{\int_0^\infty \sigma(E_n, B)E_{\bar{\nu}_e}^{2} f_{\rm FD}(E_{\bar{\nu}_e}, \eta_{\bar{\nu}_e}; T_\nu) g\Big[E_n, -\mu_e(B);T_{e}\Big] }{ \int_0^\infty E_{\bar{\nu}_e}^{4} f_{\rm FD}(E_{\bar{\nu}_e}, \eta_{\bar{\nu}_e}; T_{\bar{\nu}_e})}.\nonumber
\end{eqnarray}

\begin{figure}
\centering
\includegraphics[width=0.5\textwidth,clip=true,trim=0cm 0cm 0cm 0cm]{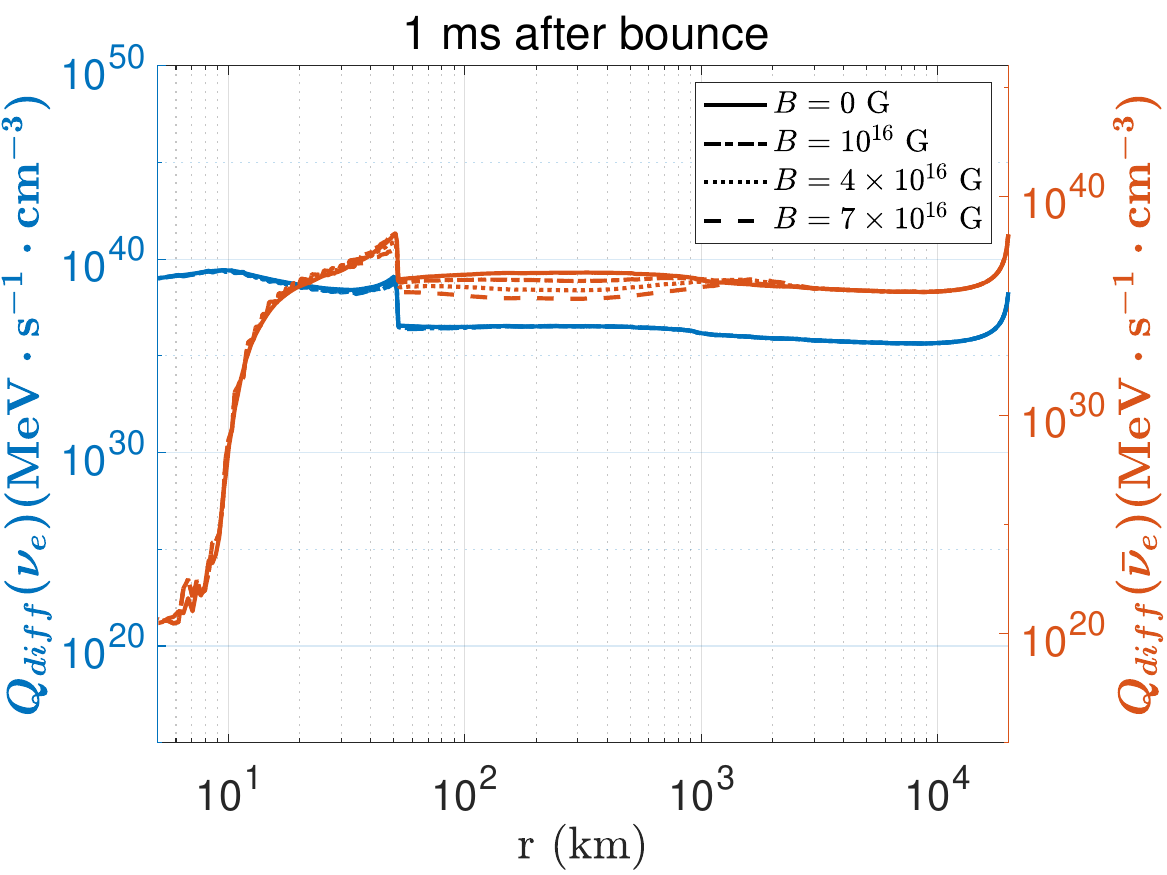}
\includegraphics[width=0.5\textwidth,clip=true,trim=0cm 0cm 0cm 0cm]{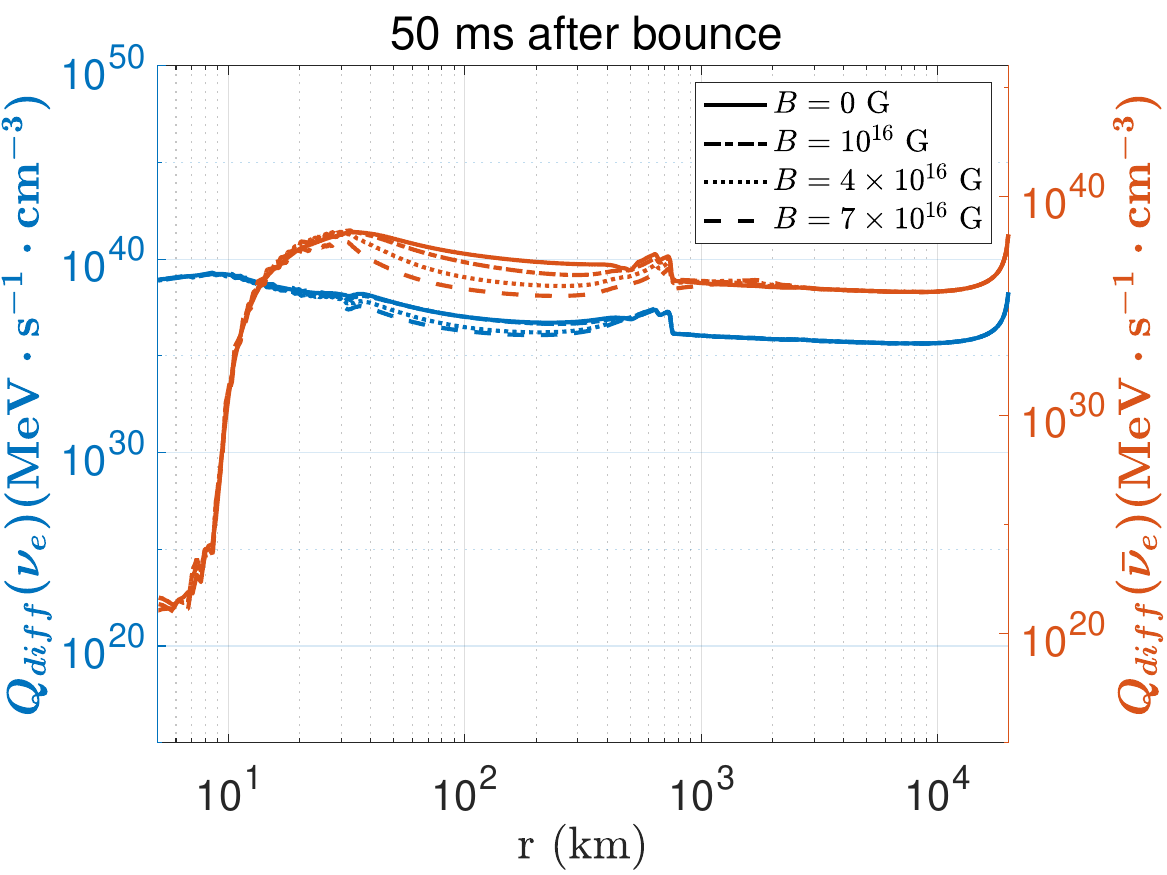}
\includegraphics[width=0.5\textwidth,clip=true,trim=0cm 0cm 0cm 0cm]{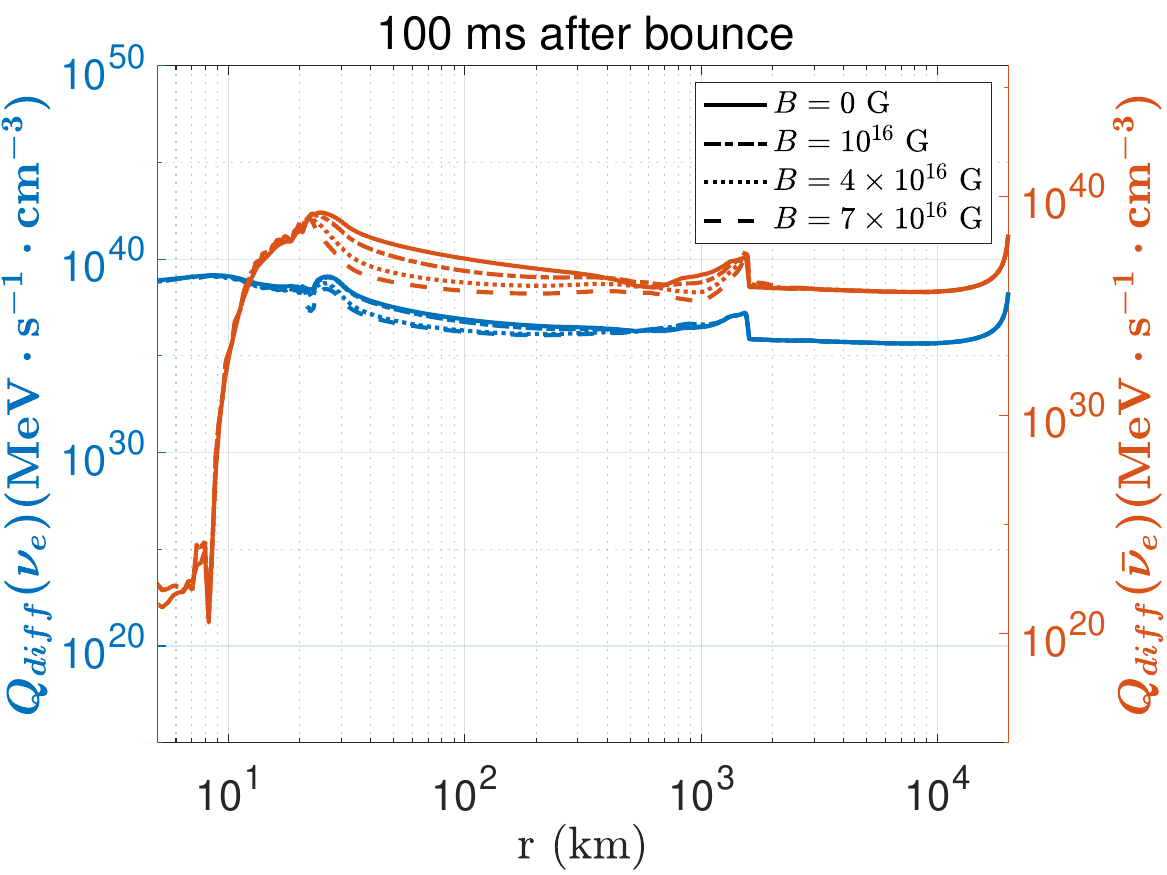}
\caption{\label{diff_compare}The diffusion rate of $\nu_e, \bar{\nu}_e$ inside the CCSN at different bouncing phases (1 ms, 50 ms and 100 ms, respectively). The blue lines are the emission rate for $\nu_e$ and the red lines are for $\bar{\nu}_e$. The results for the cases $B = 0$, $B = 10^{16}$ G, $B = 5\times 10^{16}$ G and  $B= 10^{17}$ G are shown in solid, dash-dotted, dotted, and dashed lines, respectively.}
\end{figure}

Fig. \ref{diff_compare} illustrates the local diffusion rate of neutrino.

\begin{figure}
\centering
\includegraphics[width=0.5\textwidth,clip=true,trim=0cm 0cm 0cm 0cm]{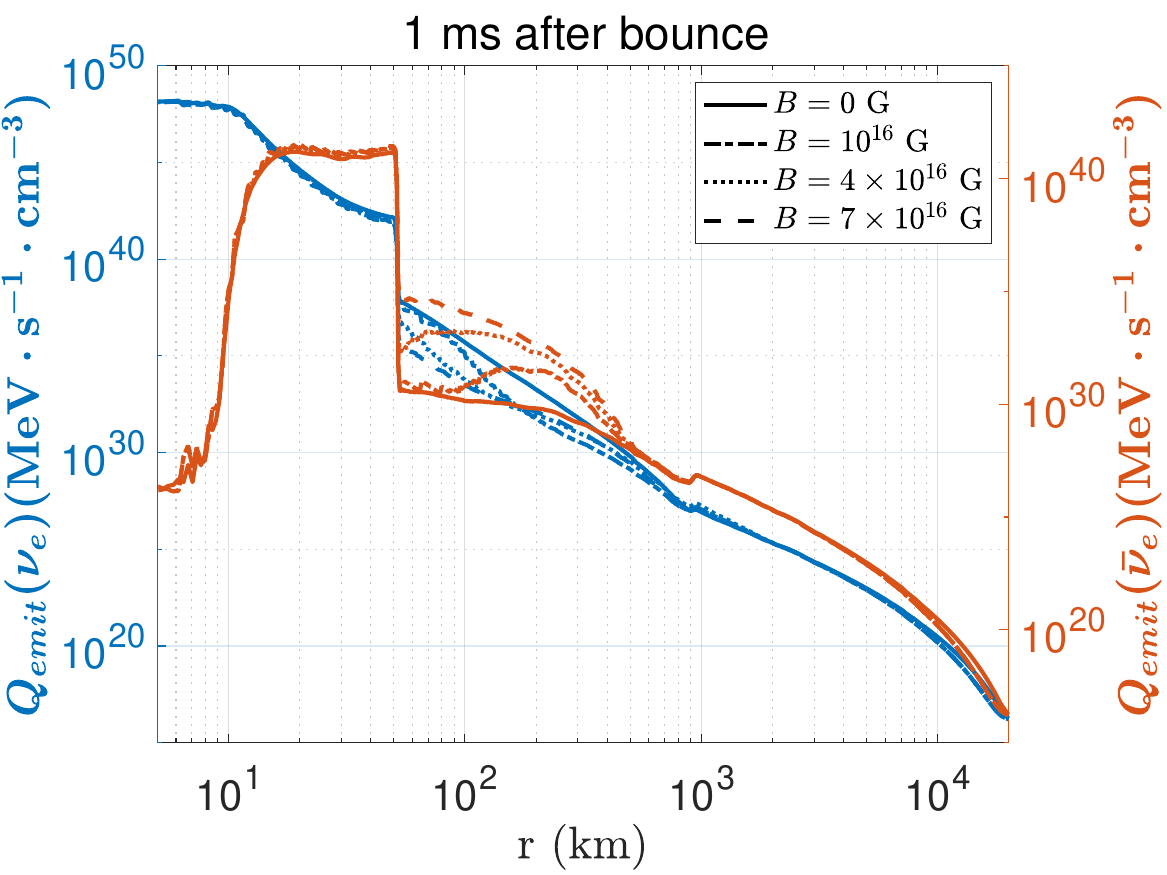}
\includegraphics[width=0.5\textwidth,clip=true,trim=0cm 0cm 0cm 0cm]{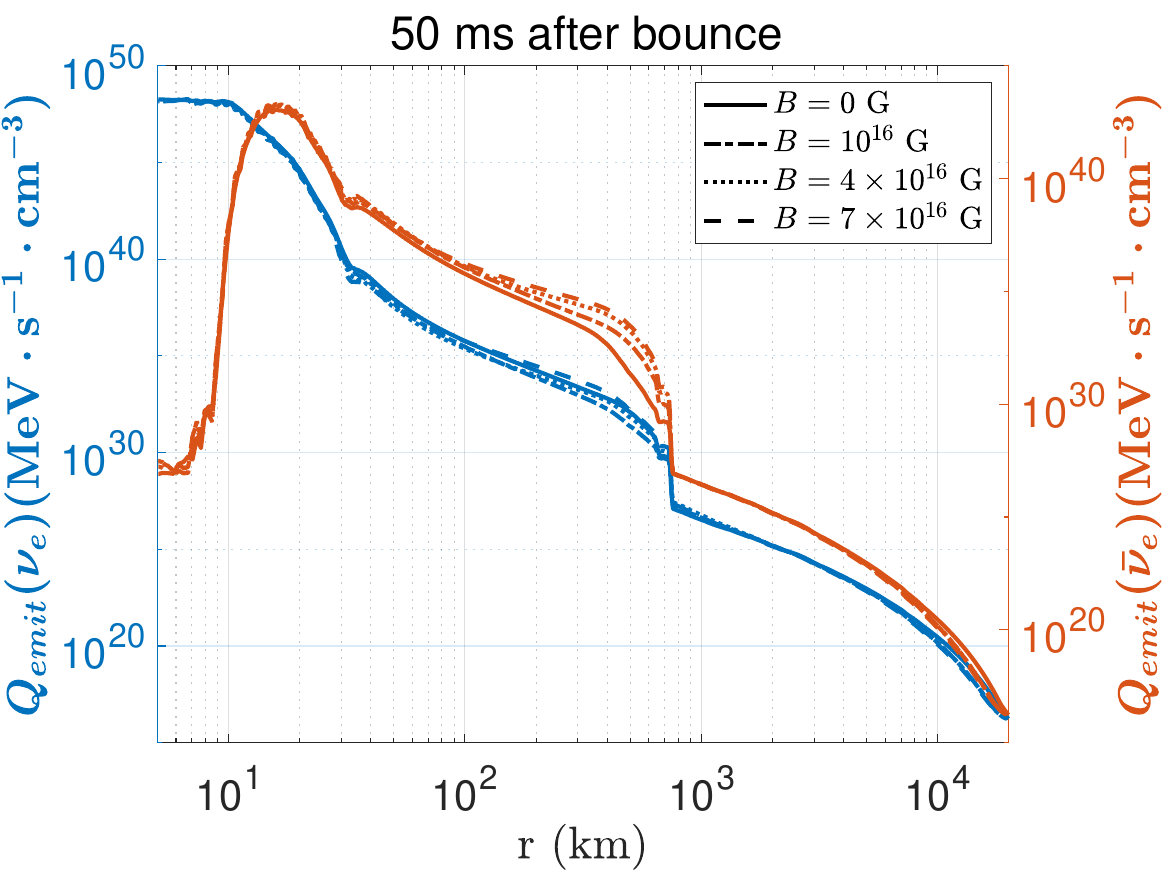}
\includegraphics[width=0.5\textwidth,clip=true,trim=0cm 0cm 0cm 0cm]{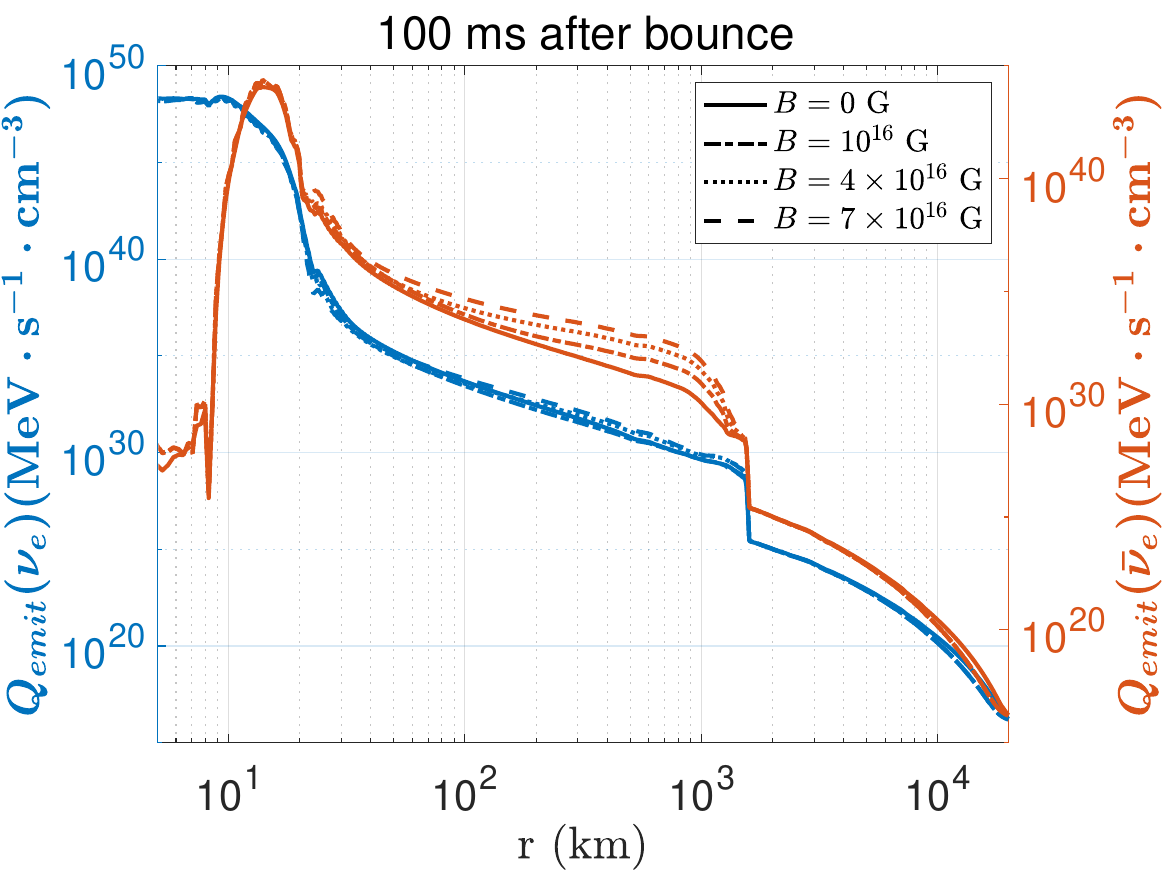}
\caption{\label{emit_compare}The emission rate of $\nu_e, \bar{\nu}_e$ inside the CCSN at different bouncing phases. The line-color convention follows Fig. \ref{diff_compare}.}
\end{figure}

Fig. \ref{emit_compare} compares the local emission rate during different phases of CCSN. 

\bibliography{bibtex}

\end{document}